\documentclass[fleqn,usenatbib]{mnras}
\usepackage[T1]{fontenc}
\usepackage{orcidlink}
\DeclareRobustCommand{\VAN}[3]{#2}
\let\VANthebibliography\thebibliography
\def\thebibliography{\DeclareRobustCommand{\VAN}[3]{##3}\VANthebibliography}

\usepackage{graphicx}   
\usepackage{amsmath}    
\usepackage{amssymb}    
\usepackage{multirow}
\usepackage{array}
\usepackage{url}
\title{New analysis of the fraction of observable nights at astronomical sites based on FengYun-2 satellite data}
\author[Wang et al.]{
Xian-Yu Wang$^{1,2}$\orcidlink{0000-0002-0376-6365},
Zhen-Yu Wu$^{1,2}$\thanks{E-mail: zywu@nao.cas.cn}\orcidlink{0000-0001-8037-1984},
Jing Liu$^{1,2}$
T. Hidayat$^{3}$
\\
$^{1}$National Astronomical Observatories, Chinese Academy of Sciences, Beijing 100101, China\\
$^{2}$University of Chinese Academy of Sciences, Beijing, 100049, China\\
$^{3}$Bosscha Observatory and Astronomy Research Division, FMIPA, Institut Teknologi Bandung, Jl. Ganesha 10, Bandung 40132, Indonesia
}
\date{Accepted 2022 February 10. Received 2022 January 25; in original form 2021 July 27}

\begin{document}
\label{firstpage}
\pagerange{\pageref{firstpage}--\pageref{lastpage}}
\maketitle

\begin{abstract}
The fraction of observable nights is an essential parameter for selecting astronomical sites. In recent years, meteorological satellite data have played an essential role in recognising and providing statistics of observable nights. We present a method to estimate the fraction of observable nights based on the FengYun-2 series of geostationary meteorological satellites and weather records of multiple astronomical sites. We have calculated the fraction of observable nights at 27 sites in Indonesia and two astronomical sites in China to validate the method. The results derived from our method show good agreement with previous works. Furthermore, we have derived the yearly distribution of the fraction of observable nights above China, which indicates the area near 40$^{\circ}$N has more observable nights than other areas in China.
\end{abstract}

\section{Introduction}\label{sec:intro}
When researchers perform search campaigns (e.g., \citealt{Erasmus2001, Sarazin2006, Ma2020}) for good astronomical sites, some main astronomical weather conditions need to be considered, including the cloud coverage, sky humidity, and atmospheric transparency and turbulence (seeing). Among them, cloud coverage is the most primary and essential parameter as it directly affects the fraction of observable nights (hereafter FONs) \citep{Graham2005, Sarazin2006, Varela2008, Aksaker2015}.

The FONs is an essential criterion used to evaluate the astronomical observing conditions of observatories and potential astronomical sites \citep{Cavazzani2011, Hidayat2012, Hotan2013, Hellemeier2018}. Statistics of the FONs can facilitate the creation and adjustment of observing schedules and inform decisions regarding updating equipment at existing observatories. The long-term monitoring of observable nights at potential astronomical sites can be a primary measurement criterion for building new observatories.

There are three main kinds of data used to quantify the FONs: data from weather stations, telescope observation logs, and meteorological satellite data. Meteorological data collected from weather stations, such as the temperature, wind speed and direction, relative humidity and air pressure, can be employed to characterise the observing conditions \citep{Lombardi2006, Lombardi2007, Lombardi2008}. Such data have the merits of extensive geographic coverage and data accumulated over decades. However, the geographical distribution of weather stations is uneven, and areas with a relatively small population may likely have a sparse distribution of weather stations, adversely affecting the evaluation of observing conditions. Telescope observation logs are essential and can robustly characterise observing conditions. Such logs are usually detailed and accurate because observers record the observing conditions of every night that astronomical observations are carried out. However, observation logs are mainly based on the specific visual inspection of sky conditions made by observers, which is sensitive to their experience. Moreover, telescope observation logs can only reflect the observing conditions of the observatories themselves, and they provide only limited information for selecting new astronomical sites. In recent years, the quality of meteorological satellite data has dramatically improved and gradually been applied to astronomical site selection (e.g., \citealt{Erasmus2001, Sarazin2006, Cavazzani2011, Aksaker2015, Aksaker2020}). Similar to ground weather station data, meteorological satellite data also have the advantages of long duration and extensive geographical coverage. Moreover, meteorological satellite data can be applied to any site on Earth, which is an excellent advantage for global-scale searches for suitable astronomical sites. However, compared with weather stations and telescope observation logs, meteorological satellite data have lower spatial and temporal resolutions, leading to inaccurate estimates of the local observing conditions of astronomical sites.

\begin{table}
\centering
\caption{Main technical properties of the VISSR-2 radiometer}
\label{tab:fy2_technical}
\begin{tabular}{cccr} 
\hline
Channel & Range ($\mu m$) & Ground Resolution at Nadir (km)\\
\hline
Infrared 1 & 10.3 $\sim$ 11.3 & 5 \\
Infrared 2 & 11.5 $\sim$ 12.5 & 5 \\
Infrared 3 & 6.3 $\sim$ 7.6 & 5 \\
Infrared 4 & 3.5 $\sim$ 4.0 & 5 \\
Visible  & 0.55 $\sim$ 0.99 &1.25\\
\hline
\end{tabular}
\end{table}

Since \citet{Erasmus2001, Erasmus2002} and \citet{Erasmus2006} proved the feasibility of using meteorology satellite data to quantify observing conditions for observatories or potential sites, researchers have sought to quantify the FONs at selected astronomical sites. \citet{yao2005} reported the early astronomical site selection of sites in Western China, which was based on Japanese Geostationary Meteorological Satellite (GMS)\footnote{\url{https://www.jma.go.jp/en/gms/}} and Geostationary Operational Environmental Satellite (GOES)\footnote{\url{https://www.goes.noaa.gov/}} data. For the site selection of the European Extremely Large Telescope (E-ELT), \citet{Sarazin2006} developed a site selection tool, FriOWL, using long-term re-analysis climate data provided by the European Centre for Medium-Range Weather Forecasts (ECMWF\footnote{\url{https://www.ecmwf.int/}}) and the geographical information system (GIS). Based on infrared images obtained with the GOES12 satellite, \citet{Cavazzani2011} proposed a threshold night classification method and applied it to Paranal, La Silla, La Palma, Mt Graham, and Tolonchar. \citet{Hidayat2012} utilized 6.7- and 10.7-$\mu$m data from GOES9 to calculate the FONs in Indonesia. Recently site selection exercises (e.g., \citealt{Aksaker2015, Daniyal2019, Aksaker2020}) adopted multi-criteria decision analysis (MCDA) based on GIS.

Recently, site testing campaign for the Large Optical/infrared Telescope (LOT) of China, a ground-based 12 m diameter telescope, is undergoing \citep[e.g.,][]{Cao2020, Feng2020}, which motivates us to search for the potential best sites in China. Up to now, almost the best sites are located in the Western Hemisphere \citep{Vernin2008, Schock2009}. China potentially provides high-quality sites for astronomical observations, with the largest geographical area in the Eastern Hemisphere.

The observable night is an essential factor in site selection. However, previous works have different definitions of it. \cite{Hidayat2012} and \cite{Erasmus2002} gathered samples of a 5 $\times$ 5-pixel region centred on the point of interest and defined an observable night when at least 20 of 25 pixels is clear. \cite{Cavazzani2011}  defined a observable night as a night when brightness temperatures ($T_{\rm B}$) of a pixel of matrix 1$^{\circ} \times 1^{\circ}$ satisfies the condition of 2$\sigma < T_{\rm B}^{\rm Max} - T_{\rm B} < 3\sigma$, where 1$\sigma$ is the night brightness temperature range and $T_{\rm B}^{\rm Max}$ is the largest $T_{\rm B}$ in a month. In this work, we considered both the astronomical observing sky and the length of observable time. We defined the astronomical observing sky area as the pixels within a 60$^{\circ}$ zenith 0n the satellite image and requested an observable night has more than three hours for astronomical observation.

The present work introduces a method that uses satellite and ground data to estimate the FONs. In this work, the satellite data of the Chinese FengYun-2 (FY-2) series are adopted, in which the cloud total amount (CTA) and upper Troposphere humidity (UTH) are utilised. This is the first time that Fengyun data are adopted to estimate FONs. The ground data include the observation logs of the 60/90-cm Schmidt telescope at Xinglong Station, and the summaries of the FONs provided by \citet{zhang2015}, \citet{Tian2016} and \citet{Xin2020}. The relatively low spatial resolution of the satellite data can adversely affect the estimates of the FONs, resulting in considerable uncertainty. However, our use of the ground data (i.e., the records of the local weather conditions of astronomical sites) helps to reduce these uncertainties. 

The remainder of this paper is organised as follows. In Sections~\ref{sec: satellite data} and~\ref{sec: ground data}, the satellite and ground data used in this work are described, respectively. The methodology is described in Section~\ref{sec: Methodology}. Section~\ref{sec: Comparative analysis and Discussion} presents comparative analysis and discussion. The conclusions are summarised in Section~\ref{sec:conclusions}.

\begin{table}
\centering
\caption{Satellite data used in this work }
\label{tab:fy2_data_detail}
\begin{tabular}{lccr}
\hline
Satellite & Central Longitude & Time Range\\
\hline
FY-2D &  86$^{\circ}$ .5 E & Jan 01, 2008 -- Dec 31, 2010 \\
FY-2E & 105$^{\circ}$  E &  Jan 01, 2011 -- Jun 02, 2015\\
FY-2G  & 105$^{\circ}$  E& Jun 03, 2015 -- Apr 12, 2018\\
FY-2G & 99$^{\circ}$ .5 E& Apr 13, 2018 -- Dec 31, 2019\\
\hline
\end{tabular}
\end{table}

\section{Satellite data}\label{sec: satellite data}

The data of meteorological satellites provide valuable information for the characterisation of astronomical sites \citep{Erasmus2002, Sarazin2006, Cavazzani2011}. There are two weather satellites by orbit type: polar-orbiting and geostationary meteorological satellites. The latter is more advantageous in searching for good sites than the former. Polar-orbiting meteorological satellites have an orbital altitude of $\sim$1,000 km and can produce images for a site every 12 hours. In contrast, geostationary meteorological satellites have a high orbital altitude ($\sim$36,000km) and can produce images for near half of the globe, even every 15 minutes.

Currently, among geostationary meteorological satellites, only GMS and FY-2 satellites can cover most areas of China over the past decades. On GMS images, China is positioned near the edge, which potentially has an adverse influence on the data quality of the area of China. Therefore, the data of the FY-2 satellite was our primary choice for searching good sites in China.

\begin{figure*}
\includegraphics[width=0.5\textwidth, trim=10 100 0 320, clip]{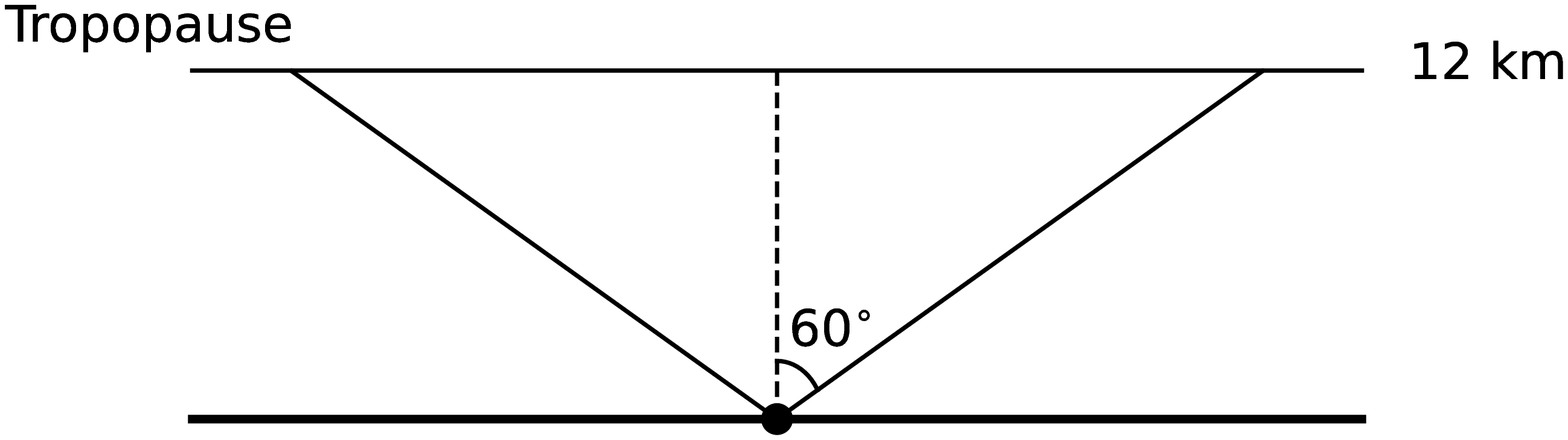}
\includegraphics[width=0.45\textwidth, trim=20 0 50 40, clip]{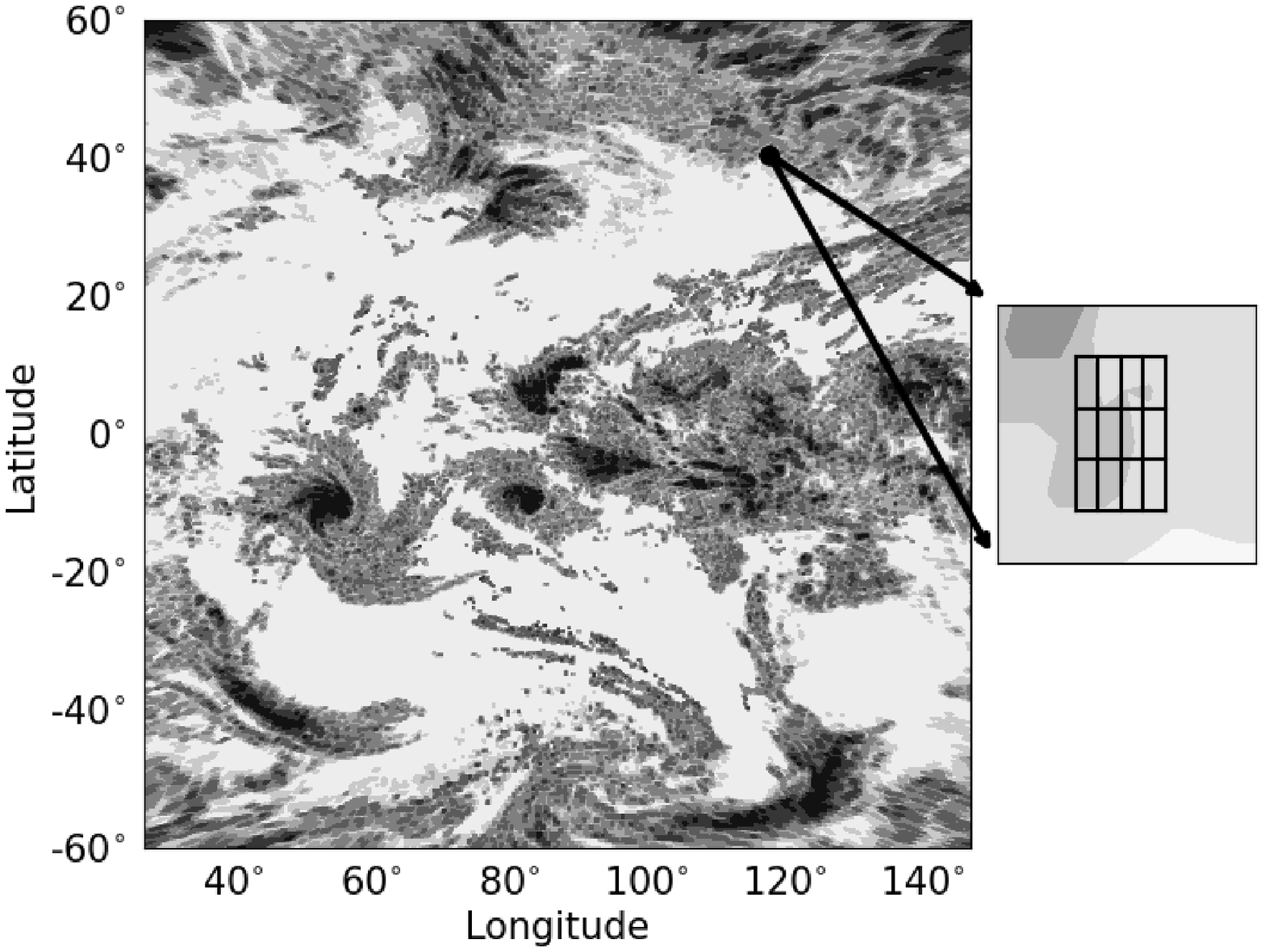}
\caption{Left picture shows a schematic diagram of observable sky area. The black dot represents the locations of current astronomical sites. The observable sky area includes the pixels corresponding the area within a 60$^{\circ}$ zenith. Right picture is an application example for Xinglong Station ($117^{\circ}34^{\prime}38\arcsec.78$E, $40^{\circ}23^{\prime}43\arcsec.98$N). For this site, the observable sky can be represented by 12 pixels, as shown in the local, enlarged drawing.}
\label{fig: skyarea}
\end{figure*}
Since the successful launch of the first of the FY-2 series of Chinese geostationary meteorological satellites, they have blossomed into a large family of eight, including FY-2A to FY-2H. Among them, FY-2A and FY-2B are experimental meteorological satellites that have not yielded any available scientific data. The geostationary operational meteorological satellites, FY-2C to FY-2H, have yielded a variety of scientific data, such as atmospheric motion vectors (AMVs), cloud detection (CLD), cloud top temperature(CTT), CTA, and UTH. These data are derived from images produced by the Stretched Visible and Infrared Spin Scan Radiometer 2 (VISSR-2), which has been adopted for all FY-2 series satellites except FY-2A. VISSR-2 has five channels, including one visible band and four infrared bands. The main technical parameters of VISSR-2 are shown in Table~\ref{tab:fy2_technical}. To date, the accumulated images span 15 years, which provide the potential to study the FONs to facilitate the selection of astronomical sites.

In this work, data obtained by FY-2D, FY-2E, and FY-2G are used because their images can cover more areas of China than images obtained by the other satellites in the FY-2 series. Among the various products of the FY-2 satellites, CTA and UTH have been proven to have a close relation with observable nights \citep{Erasmus2002, Aksaker2015}. Therefore, we chose them for the period 2008--2019 to calculate the FONs, as shown in Table~\ref{tab:fy2_data_detail}. Note that the data for 2013 were not used due to the incompleteness of this year. Data obtained by the FY-2 series satellites are available at the website of the National Satellite Meteorological Center (NSMC)\footnote{\url{http://www.nsmc.org.cn/nsmc/en/home/index.html}}, China Meteorological Administration.

\subsection{Cloud total amount -- CTA}\label{Dataset_CTA}

According to \citet{glickman2000}, cloud coverage or cloudiness, i.e. CTA, refers to the percentage of the sky covered with clouds at the desired geographical coordinates. Astronomical observations require crucial weather conditions, and CTA is essential for astronomical site selection and evaluation \citep{Graham2005, Sarazin2006, Varela2008, Aksaker2015, Aksaker2020}. As one of the most important basic products of the FY-2 series satellites, CTA can be derived from images obtained in the visible and infrared channels. Following \citet{Xu2008}, we present the basic principle of CTA as follows. All radiation received by a radiometer can be represented as:

\begin{equation}
I=\left(1-A_{c}\right) I_{\mathrm{clr}}+A_{c} I_{\mathrm{cld}}
\end{equation}

where $A_{c}$ is the CTA, $I_{\rm cld}$ is the radiation in a cloudy pixel, and $I_{\rm clr}$ is the radiation in a clear pixel. Then, CTA can be calculated by:

\begin{equation}
A_{c}=\left(I_{\mathrm{clr}}-I\right) /\left(I_{\mathrm{clr}}-I_{\mathrm{cld}}\right)
\end{equation}

The CTA used in this work has a spatial resolution of 5 km at nadir and a temporal resolution of one hour.

\subsection{Upper Troposphere humidity - UTH}\label{Dataset_UTH}

The internal relationship between UTH and weather conditions was first reported by \citet{Erasmus2002}. Since then, UTH has become essential data for evaluating astronomical sites. 

For the FY-2 series satellites, UTH is the average relative humidity of the atmosphere from 3000 m to 7000 m  \citep{Xu2008}. The spatial and temporal resolutions of UTH are 5 km at nadir and three hours, respectively. Moreover, to ensure data accuracy, only data within the satellites zenith of 75$^{\circ}$ are adopted. Therefore, a small area of north-eastern China has no data.

\begin{table*}
\caption{Monthly statistics of the observable nights (given as a percentage) for Xinglong Station, Delingha Station and Lijiang Station derived from ground data}
\begin{tabular}{lccccccccccccc}
\hline
\multicolumn{13}{c}{The FONs at Xinglong Station derived from \citet{zhang2015}}\\
\hline
& Jan    & Feb    & Mar   & Apr   & May   & Jun   & Jul   & Aug   & Sep   & Oct   & Nov    & Dec   \\
& ($\%$)    & ($\%$)    & ($\%$)   & ($\%$)   & ($\%$)   & ($\%$)   & ($\%$)   & ($\%$)   & ($\%$)   & ($\%$)   & ($\%$)    & ($\%$)   \\
\hline
2008 & 87.10  & 96.55  & 45.16 & 50.00 & 35.48 & 33.33 & 29.03 & 32.26 & 50.00 & 83.87 & 93.33  & 90.32 \\
2009 & 80.65  & 64.29  & 51.61 & 56.67 & 64.52 & 66.67 & 35.48 & 35.48 & 50.00 & 74.19 & 70.00  & 70.97 \\
2010 & 77.42  & 53.57  & 41.94 & 40.00 & 54.84 & 36.67 & 25.81 & 29.03 & 63.33 & 67.74 & 100.00 & 93.55 \\
2011 & 100.00 & 67.86  & 64.52 & 66.67 & 58.06 & 50.00 & 35.48 & 35.48 & 60.00 & 74.19 & 60.00  & 83.87 \\
2012 & 80.65  & 100.00 & 70.97 & 53.33 & 64.52 & 46.67 & 22.58 & 38.71 & 73.33 & 87.10 & 80.00  & 67.74 \\
2014 & 93.55  & 53.57  & 87.10 & 73.33 & 58.06 & 53.33 & 22.58 & 51.61 & 50.00 & 58.06 & 90.00  & 90.32 \\
\hline
\multicolumn{13}{c}{The FONs at Xinglong Station derived from the observation logs of the 60/90-cm Schmidt Telescope}\\
\hline
     & Jan    & Feb    & Mar   & Apr   & May   & Jun   & Jul   & Aug   & Sep   & Oct   & Nov    & Dec   \\
     & ($\%$)    & ($\%$)    & ($\%$)   & ($\%$)   & ($\%$)   & ($\%$)   & ($\%$)   & ($\%$)   & ($\%$)   & ($\%$)   & ($\%$)    & ($\%$)   \\
\hline
2015 & 83.87  & -    & 77.42 & 33.33 & 54.84 & -   & 35.48 & 25.81 & -   & -   & -    & -   \\
2016 & -    & -    & -   & 33.33 & 38.71 & -   & 19.35 & 6.45  & 33.33 & -   & 46.67  & 80.65 \\
2017 & -    & -    & 64.52 & -   & 64.52 & 43.33 & 6.45  & 0.00  & -   & 45.16 & 93.33  & 90.32 \\
2018 & 87.10  & -    & 51.61 & 53.33 & 45.16 & -   & 3.23  & 6.45  & -   & -   & -    & 51.61 \\
\hline
\multicolumn{13}{c}{The FONs at Delingha Station derived from \citet{Tian2016}}\\
\hline
& Jan    & Feb    & Mar   & Apr   & May   & Jun   & Jul   & Aug   & Sep   & Oct   & Nov    & Dec   \\
& ($\%$)    & ($\%$)    & ($\%$)   & ($\%$)   & ($\%$)   & ($\%$)   & ($\%$)   & ($\%$)   & ($\%$)   & ($\%$)   & ($\%$)    & ($\%$)   \\
\hline
2011 & 61.29  & 75.00  & 58.06 & 83.33 & 58.06 & 53.33 & 54.84 & 58.06 & 73.33 & 87.10 & 73.33  & 93.55 \\
2012 & 77.42  & 72.41  & 74.19 & 66.67 & 41.94 & 40.00 & 29.03 & 41.94 & 83.33 & 90.32 & 60.00  & 64.52 \\
2014 & 80.65  & 64.29  & 74.19 & 60.00 & 77.42 & 60.00 & 64.52 & 61.29 & 63.33 & 74.19 & 80.00  & 87.10 \\
\multicolumn{13}{c}{The FONs at Lijiang Station derived from \citet{Xin2020}}\\
\hline
& Jan    & Feb    & Mar   & Apr   & May   & Jun   & Jul   & Aug   & Sep   & Oct   & Nov    & Dec   \\
& ($\%$)    & ($\%$)    & ($\%$)   & ($\%$)   & ($\%$)   & ($\%$)   & ($\%$)   & ($\%$)   & ($\%$)   & ($\%$)   & ($\%$)    & ($\%$)   \\
\hline
2019 & 96.77  & 75.00  & 77.42 & 66.67 & 61.29 & 26.67 & 3.23  & 35.48 & 13.33 & 54.84 & 86.67  & 96.77 \\
\hline
\end{tabular}
\label{tab: ground data}
\begin{flushleft}
\textbf{Note:} `-' means a record in which more than two days are missing for the month.  \\
\end{flushleft}
\end{table*}

\section{Ground data}\label{sec: ground data}

In this section, we present the sources of the ground data, including the observable night records from the observation logs of the 60/90-cm Schmidt telescope at Xinglong Station, and the astronomical observing conditions summaries at Xinglong Station, Denglingha Station and Lijiang Station conducted by \citet{zhang2015}, \citet{Tian2016} and \citet{Xin2020}, respectively. Based on the ground data, we summarised the monthly statistics of the observable nights for these three sites in Table~\ref{tab: ground data}. 

\subsection{Xinglong Station}

Xinglong Station of the National Astronomical Observatories, Chinese Academy of Sciences (NAOC), is an optical astronomical observatory in China. It is 120 km from Beijing and has an altitude of 960 m. This work uses two kinds of data related to the number of observable nights at Xinglong Station. The first, presented by \citet{zhang2015}, consists of a summary of observable nights at Xinglong Station from 2007 to 2014 based on observation logs from the 2.16-m telescope, 80-cm telescope, and 85-cm telescope. The second is the observation logs of the 60/90-cm Schmidt telescope, which covers 2015--2018. Based on the work of \citet{zhang2015}, we considered an observable night as a night when observations can be performed in both clear sky and partly cloudy conditions. 
The 60/90-cm Schmidt telescope observation logs were incomplete due to downtime, bad weather, and other reasons. We employed these data from 2015 to 2018 to extend the time coverage of data for Xinglong Station. For the 60/90-cm Schmidt telescope, an observable night was considered as a night with at least three hours of astronomical observations. Moreover, to derive an accurate estimate of the FONs for each month, we did not use data from any month in which the records had more than two days of missing logs.

\subsection{Delingha Station}
Delingha Station is located 40 km east of the city of Delingha, which is situated on the Qinghai--Tibetan Plateau. \cite{Tian2016} presented statistics of the observable nights at Delingha Station from 2011 to 2014. According to \cite{Tian2016}, there are three kinds of nights: clear, cloudy, and useless nights. A clear night means the night is clear for at least six hours \citep{Morrison1973} and a cloudy night refers to a night with 65$\%$ or less sky coverage \citep{Tapia1992}. For Delingha Station, the number of observable nights was considered as the sum of clear nights and cloudy nights presented by \citet{Tian2016}.

\subsection{Lijiang Station}
Lijiang Station, an optical astronomical site in South China, is 40 km from the city of Lijiang. \citet{Xin2020} introduced the Astronomical Site Monitoring System (ASMS) installed at Lijiang Station and presented its observing conditions. According to \citet{Xin2020}, a photometric time block (PTB) is described as a one-hour period where the cloud cover is $\leq$ 30$\%$ and 10 minutes of interruption time is allowed. A spectroscopic time block (STB) has a similar definition, where the only difference is the maximum cloud cover, which is 50$\%$. Moreover, a photometric night is defined as a night when the sum of PTB exceeds three hours, and a spectroscopic (observable) night is defined as a night when the sum of STB exceeds three hours. For Lijiang Station, we considered the FONs as the fraction of spectroscopic nights given by \citet{Xin2020}.

\begin{table*}
\caption{Yearly FONs from 2008 to 2010 for 27 sites in Indonesia}
\label{tab: long-term comparison}
\begin{tabular}{c|ccc|ccc|cccc}
\hline
Site      & \multicolumn{3}{c|}{\citet{Hidayat2012}}     & \multicolumn{3}{c|}{This Work}        & \multicolumn{4}{c}{\citet{Hidayat2012} - This work} \\
\cline{2-4} \cline{5-7} \cline{8-11}
              &2008&2009&2010 &2008&2009&2010&  2008&2009&2010&Average\\
               &  ($\%$) & ($\%$) & ($\%$)      & ($\%$)     & ($\%$)& ($\%$)& ($\%$)  & ($\%$)& ($\%$)        &($\%$)         \\ \hline
Alor             	&	65.10 	&	65.30 	&	57.60 	&	52.02 	&	65.23 	&	48.73 	&	13.08 	&	0.07 	  &	8.87  	&	7.34 	\\
Binohoe          	&	42.80 	&	51.00 	&	50.50 	&	17.73 	&	46.89 	&	20.40 	&	25.07 	&	4.11 	  &	30.10 	&	19.76 	\\
E. Sumbawa       	&	67.00 	&	66.80 	&	59.00 	&	57.70 	&	72.48 	&	49.90 	&	9.30 	  &	-5.68 	&	9.10 	  &	4.24 	\\
Erekebo          	&	9.60 	  &	14.00 	&	16.90 	&	6.37 	  &	19.23 	&	9.45 	  &	3.23 	  &	-5.23 	&	7.45 	  &	1.82 	\\
Kerinci          	&	32.30 	&	32.80 	&	32.00 	&	22.11 	&	35.11 	&	8.35 	  &	10.19 	&	-2.31 	&	23.65  	&	10.51 	\\
Kupang           	&	67.90 	&	69.80 	&	64.50 	&	58.80 	&	70.43 	&	55.85 	&	9.10 	  &	-0.63 	&	8.65 	  &	5.71 	\\
Lembang          	&	44.10 	&	45.70 	&	24.70 	&	44.90 	&	53.18 	&	23.00 	&	-0.80 	&	-7.48 	&	1.70 	  &	-2.19 	\\
Lembata          	&	63.90 	&	64.50 	&	59.50 	&	54.00 	&	67.97 	&	48.60 	&	9.90   	&	-3.47 	&	10.90 	&	5.78 	\\
Lombosang        	&	43.70 	&	51.70 	&	36.10 	&	34.50 	&	52.09 	&	18.07 	&	9.20   	&	-0.39 	&	18.03 	&	8.95 	\\
Mutis            	&	63.80 	&	65.70 	&	58.90 	&	56.54 	&	70.36 	&	54.35 	&	7.26   	&	-4.66 	&	4.55 	  &	2.38 	\\
P. Jaya 1        	&	4.60 	  &	6.60 	  &	9.70 	  &	7.12 	  &	22.66 	&	10.54 	&	-2.52 	&	-16.06 	&	-0.84 	&	-6.47 	\\
P. Jaya 2        	&	11.00 	&	13.90 	&	19.70 	&	6.09 	  &	19.99 	&	13.35 	&	4.91   	&	-6.09 	&	6.35 	  &	1.72 	\\
Rantemario       	&	25.10 	&	35.40 	&	20.00 	&	21.70 	&	42.71 	&	11.36 	&	3.40   	&	-7.31 	&	8.64 	  &	1.58 	\\
Rinjani 1        	&	63.20 	&	65.10 	&	56.60 	&	57.43 	&	68.38 	&	49.28 	&	5.77   	&	-3.28 	&	7.32 	  &	3.27 	\\
Rinjani 2        	&	64.10 	&	65.90 	&	57.20 	&	56.88 	&	69.40 	&	49.90 	& 7.22   	&	-3.50 	&	7.30 	  &	3.67 	\\
Ruteng           	&	65.20 	&	64.40 	&	58.10 	&	59.41 	&	70.77 	&	50.72 	&	5.79   	&	-6.37 	&	7.38 	  &	2.27 	\\
S. Waingapu      	&	68.20 	&	69.50 	&	66.90 	&	58.80 	&	72.01 	&	50.10 	&	9.40  	&	-2.51 	&	16.80 	&	7.90 	\\
Sangihe-Siau     	&	46.70 	&	53.70 	&	55.50 	&	24.57 	&	46.00 	&	46.54 	&	22.13 	&	7.70 	  &	8.96 	  &	12.93 	\\
Sawu             	&	72.20 	&	71.40 	&	71.80 	&	61.67 	&	71.39 	&	59.89 	&	10.53 	&	0.01 	  &	11.91 	&	7.48 	\\
Sibayak          	&	15.00 	&	15.30 	&	15.70 	&	10.47 	&	19.64 	&	7.80 	  &	4.53  	&	-4.34 	&	7.90   	&	2.70 	\\
Silimapuluh      	&	15.50 	&	16.50 	&	16.20 	&	11.50 	&	20.81 	&	5.54 	  &	4.00  	&	-4.31 	&	10.66 	&	3.45 	\\
Sinabung         	&	16.20 	&	15.20 	&	16.70 	&	10.13 	&	19.85 	&	7.73 	  &	6.07  	&	-4.65 	&	8.97 	  &	3.46 	\\
Sirung           	&	63.10 	&	65.40 	&	59.70 	&	53.32 	&	67.08 	&	48.73 	&	9.78 	  &	-1.68 	&	10.97 	&	6.36 	\\
Tibo             	&	23.10 	&	32.30 	&	20.30 	&	20.05 	&	41.75 	&	10.54 	&	3.05 	  &	-9.45 	&	9.76   	&	1.12 	\\
Timau            	&	65.20 	&	66.30 	&	59.80 	&	56.40 	&	70.09 	&	55.03 	&	8.80 	  &	-3.79 	&	4.77  	&	3.26 	\\
West Ijen        	&	58.70 	&	60.60 	&	48.70 	&	53.59 	&	64.89 	&	43.12 	&	5.11 	  &	-4.29 	&	5.58   	&	2.13 	\\
Wetar            	&	64.30 	&	63.20 	&	56.90 	&	50.99 	&	62.97 	&	48.19 	&	13.31 	&	0.23 	  &	8.71  	&	7.42 	\\
\hline
\end{tabular}
\end{table*}

\section{Methodology} \label{sec: Methodology}

In this work, we employed a threshold method to recognise observable nights. Previous works have explored many possibilities to recognise observable nights using threshold methods (e.g., \citealt{Cavazzani2011, Hidayat2012}). However, there is no standard method to choose thresholds. Based on the distribution of monthly brightness temperatures, \citet{Cavazzani2011} put forward a monthly brightness temperature threshold to recognise observable nights. \citet{Hidayat2012} calculated three different brightness temperature thresholds and chose the best one based on the $\chi^2$ analysis. We minimised the difference between computational results based on satellite and ground-based data to find the best thresholds. We created a grid of thresholds of CTA and UTH and calculated the differences between computational results and ground-based data for each pair of thresholds. The thresholds for CTA and UTH corresponding to the smallest difference were selected as the best threshold.

For each astronomical site, calculation of the astronomical observing sky area is necessary. We defined the astronomical observing sky as the sky area that a telescope could observe, corresponding to a group of pixels 0n the satellite images. \citet{Erasmus2002} used a nine-pixel area centred on astronomical sites to represent the astronomical observing sky. \citet{Cavazzani2011} introduced a method of using an average matrix to reduce the noise of satellite images and recognise observable nights. In our work, the mean value of the image matrix corresponding to the observable sky was used rather than the value of the single pixel of the astronomical site. We also required the angle between the observing orientation and zenith to be smaller than 60$^{\circ}$. Considering that most clouds are generated below the tropopause, the altitude of the tropopause, approximately 12 km, was adopted as the altitude of the highest cloud. Then, we used the Haversine formula \citep{Sinnott1984} to calculate the radius of the astronomical observable sky area visible through the tropopause. A pixel was included when the distance between the astronomical site and its geographical location was smaller than the radius of the observable sky area. A schematic diagram and application example of Xinglong Station are shown in Figure~\ref{fig: skyarea}.

Ground data can reveal local weather conditions at astronomical sites, which can be used to select the optimal satellite data thresholds. CTA and UTH have temporal resolutions of one and three hours, respectively. For three-hour periods, the satellites can produce four CTA data points (e.g., at 21:00, 22:00, 23:00 and 24:00) and two UTH data points (e.g., at 21:00 and 24:00). We used these data to represent the weather conditions between 21:00 to 24:00. We defined an observable night as a night when the CTA and UTH values were lower than their thresholds for three or more hours.

\begin{figure}
\includegraphics[width=0.48\textwidth]{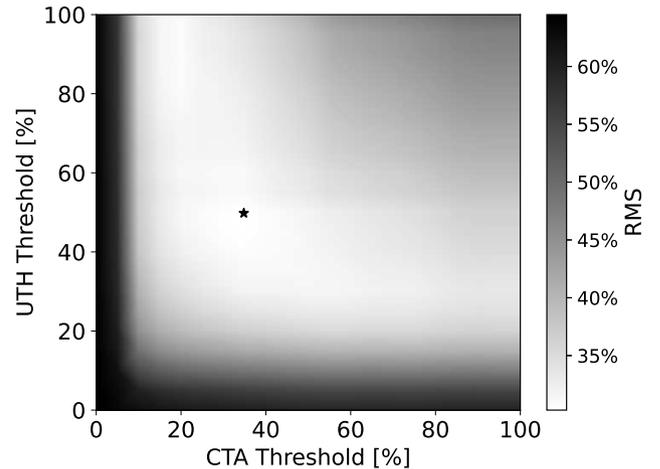}
\caption{Weighted average RMS distribution of the FONs. The black star represents the CTA and UTH thresholds with the lowest RMS.}
\label{fig: mean search Spec}
\end{figure}

To find the optimal thresholds of CTA and UTH, a direct and brute-force method was employed. We created a grid of CTA and UTH thresholds where each was allowed to vary from 0$\%$ to 100$\%$ in steps of 1$\%$. Each grid corresponded to a pair of CTA and UTH thresholds. For each grid, the deviation between the monthly FONs derived from the satellite and ground data was quantified using the root mean square (RMS) of the difference between them. Herein, we defined the monthly FONs as the fraction of the observable nights in a month, which is the same as that in \cite{zhang2015}, \cite{Tian2016}, and \cite{Xin2020}. To derive the best thresholds of CTA and UTH, we calculated the average RMS distribution (see Figure~\ref{fig: mean search Spec}) for observable nights. The best thresholds should work well for different sites that have different characteristics of weather conditions. Thus, the weights of ground data from each site are equal. According to the average RMS distribution, the optimal CTA and UTH thresholds for observable nights are 35$\%$ and 50$\%$, respectively.

\section{Comparative analysis and Discussion}
\label{sec: Comparative analysis and Discussion}
\begin{figure}
\centering
\includegraphics[width=0.48\textwidth]{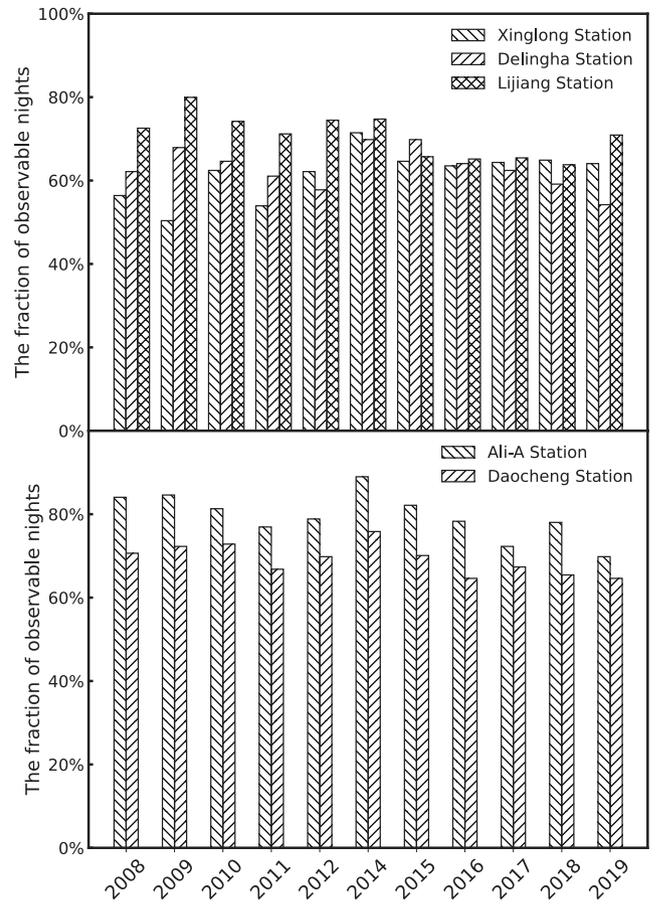}
\caption{$Upper \ panel:$ Yearly average FONs at Xinglong Station, Delingha Station and Lijiang Station. $Lower \ panel:$ Same as upper panel but for Ali-A Station and Daocheng Station.}
\label{FiveFONs}
\end{figure}

Based on the optimal thresholds of CTA and UTH, we calculated the FONs at Xinglong Station, Delingha Station, and Lijiang Station from 2008 to 2019 (except 2013). Based on the satellite data, the average FONs are 62.84$\%$, 64.21$\%$ and 71.27 $\%$, respectively (see Figure~\ref{FiveFONs}); according to ground data shown in Table~\ref{tab: ground data}, the average FONs are 62.97$\%$, 67.92$\%$ and 57.81$\%$, respectively. Note that we took the mean of the yearly average FONs from \citet{zhang2015} as the average FONs at Xinglong Station because of the incompleteness of observation logs of the 60/90-cm Schmidt telescope.

For Xinglong Station and Delingha Station, the results from the ground and satellite data are consistent within 5$\%$, but the results for Lijiang Station show a relatively large deviation (13.46$\%$). \citet{Site1999} adopted a looser definition of observable nights: nights with more than or equal to two continuous hours when cloud coverage is less than 30$\%$. Based on the observation logs of observers (Jul 1994 to Jul 1997) and weather station data (1959--1994), \citet{Site1999} predicted the long-term average of the yearly FONs at Gaomeigu in Lijiang county, namely at the location of Lijiang Station, finding 68.55$\%$, which is consistent with our estimate based on satellite data. Moreover, according to \citet{Xin2020}, the observable nights at Lijiang Station during 2019 were less than other years. Therefore, the deviation between estimates of the average FONs derived from the long-term ground and satellite data is likely smaller than the 13.46$\%$ found here.

\begin{figure*}
\centering
{\includegraphics[width=0.8\textwidth]{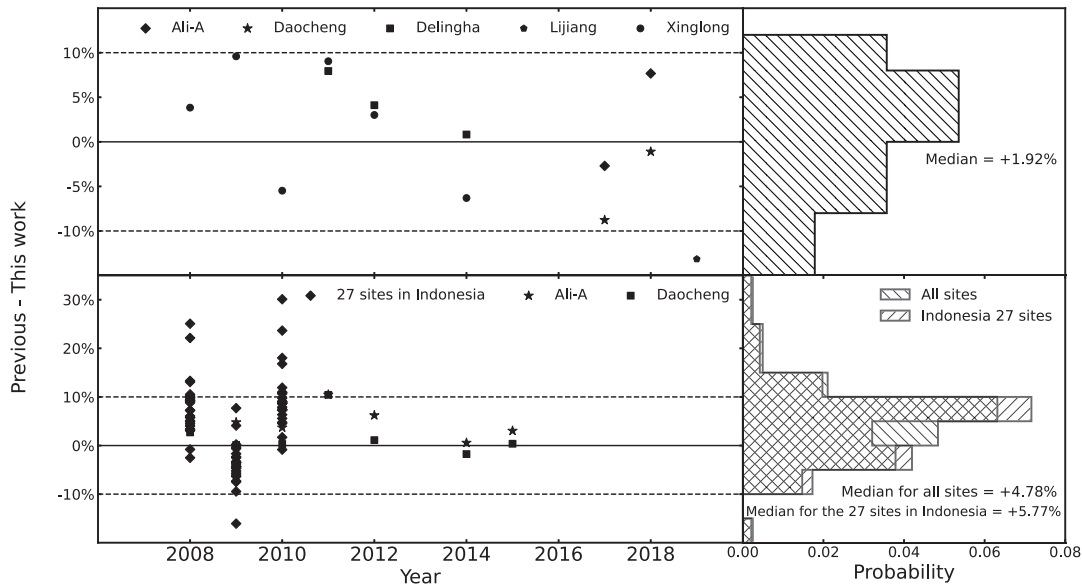}}
\caption{The differences between previous works and this work. $Upper \ panel$: FONs derived from the ground data for five sites minus our FONs estimate based on satellite data, along with it probability distribution. The median of them is +1.92$\%$.
$Lower \ panel$: FONs derived from other satellite data (27 sites in Indonesia: \citealt{Hidayat2012}; Ali-A and Daocheng: \citealt{Cao2020}) minus our results. The median of differences for all sites is +4.78$\%$, for 27 sites in Indonesia, it is +5.77$\%$.}
\label{comparsion}
\end{figure*}

We also performed an analysis of observable nights for 27 sites in Indonesia and compared our results with those of \citet{Hidayat2012}. \citet{Hidayat2012} reported the fraction of photometric and spectroscopic nights over Indonesia based on meteorological satellite data spanning 15 years (from 1996 to 2010) obtained with the Geostationary Meteorological Satellite 5 (GMS5), Geostationary Operational Environmental Satellite 9 (GOES9), and Multi-functional Transport Satellite-1R (MTSAT-1R). \citet{Hidayat2012} took an area of 5 $\times$ 5 pixels as the observable sky and assumed a pixel whose value was lower than a derived threshold as a clear pixel. \citet{Hidayat2012} adopted the definition of observable nights used by \citet{Erasmus2002}: more than 20 clear pixels indicated an observable night. Moreover, \citet{Hidayat2012} presented the mean yearly FONs and errors of 27 sites based on the 15 years of data. We applied our method to the same sites and derived the FONs based on the FY-2 data from 2008 to 2019; the results are presented in Table~\ref{tab: long-term comparison}. According to Table~\ref{tab: long-term comparison}, 51 out of 54 FONs for 27 sites of 2008 and 2010 from this work are smaller than \cite{Hidayat2012}, whereas only 5 out of 27 monthly FONs are smaller in 2009. However, 81$\%$ of absolute differences of the yearly FONs of the 27 sites for the overlapped time (2008 to 2010) between \cite{Hidayat2012} and this work are smaller than 10$\%$, and for the average FONs from 2008 to 2010, it is 89$\%$. The median of differences for 27 sites in Indonesia is +5.77$\%$ (see the lower panel of Figure~\ref{comparsion}). Therefore we conclude that: 1) there is a systemic effect due to the inherent difference between the models \cite{Hidayat2012} and we used. 2) the results from \cite{Hidayat2012} and this work are consistent.

Furthermore, our method was applied to two astronomical sites in China: Ali-A and Daocheng. Previous works calculated the FONs using different data. \citet{Cao2020} used meteorological data from the satellite of National Oceanic and Atmospheric Administration (NOAA) and GMS to analyze cloud coverage and employed data from the Moderate Resolution Imaging Spectroradiometer (MODIS) on EOS/TERRA and EOS/AQUA to calculate the annual average percentage of observable nights. Based on the cloudiness estimated by data obtained with the all-sky cameras installed at Ali and Daocheng, \citet{Feng2020} reported the annual average FONs from 2017 March 10 to 2018 March 10. The 
FONs from this work, \citet{Cao2020} and the ground data from \citet{Feng2020} are shown in Table~\ref{tab: diff}. For Ali-A and Daocheng, 89$\%$ of the absolute differences between the results from our work and from \citet{Cao2020} and \citet{Feng2020} are less than 10$\%$, which indicates good agreements. The yearly FONs of these two sites are presented in Figure~\ref{FiveFONs}.

Moreover, to check whether there is a trend in our estimate, we compared our estimate with results derived from ground data and other satellites, which is shown in the upper panel of Figure~\ref{comparsion}. The differences between our estimate from FY satellite data and ground data are within 10$\%$, indicating no significant trend. However, our estimate is slightly lower compared with results from other satellites. The median of the difference is 4.78$\%$, but if we remove the data of 2009, it becomes 7.42$\%$. 94$\%$ of differences are positive, and 78$\%$ of them are within 10$\%$. Therefore, the two comparisons indicate our estimate agrees with results from the ground data within 10$\%$ of the error and is slightly lower than results from other satellite data.

\begin{figure*}
\centering
{\includegraphics[width=0.32\textwidth]{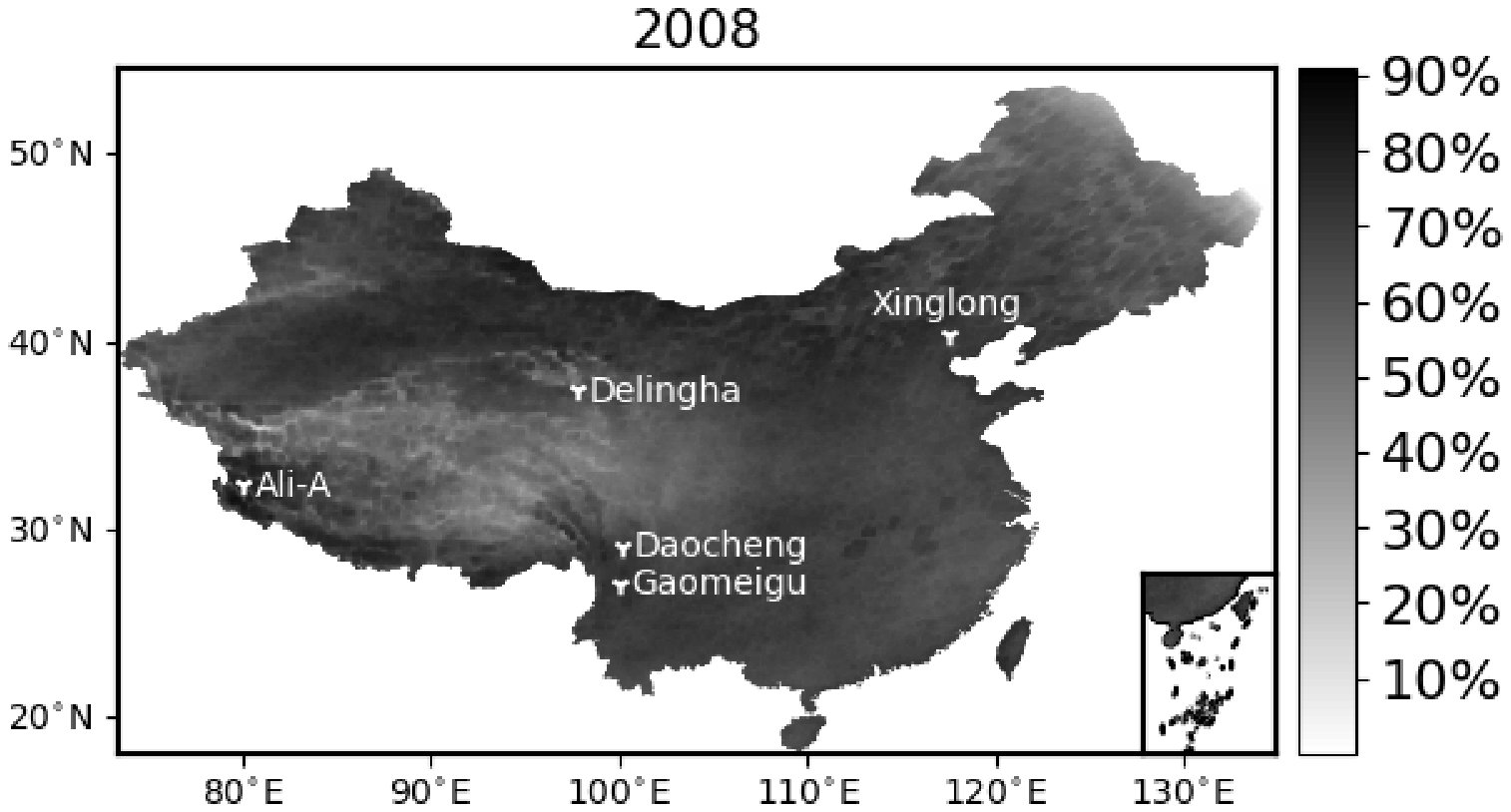}}
{\includegraphics[width=0.32\textwidth]{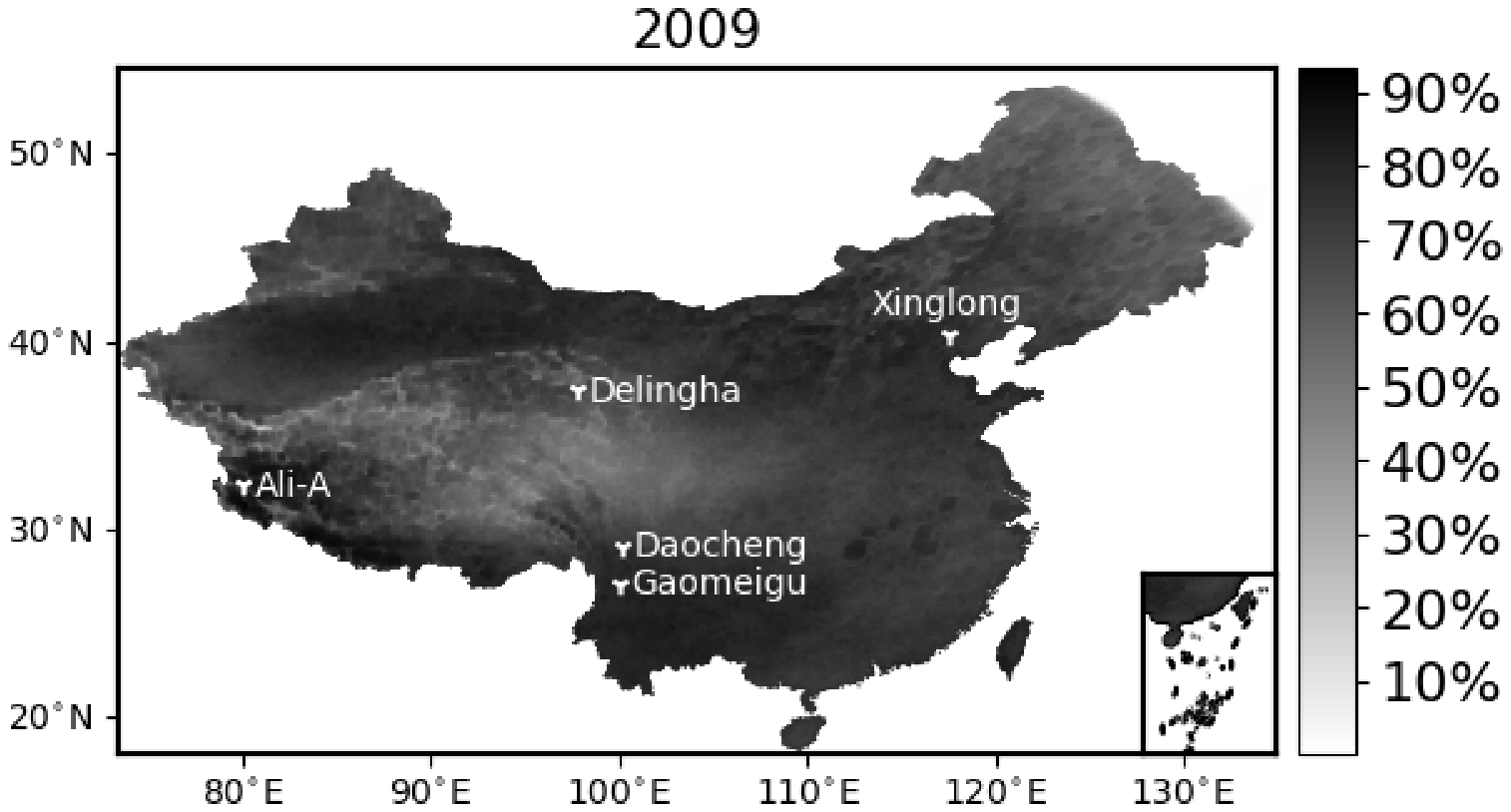}}
{\includegraphics[width=0.32\textwidth]{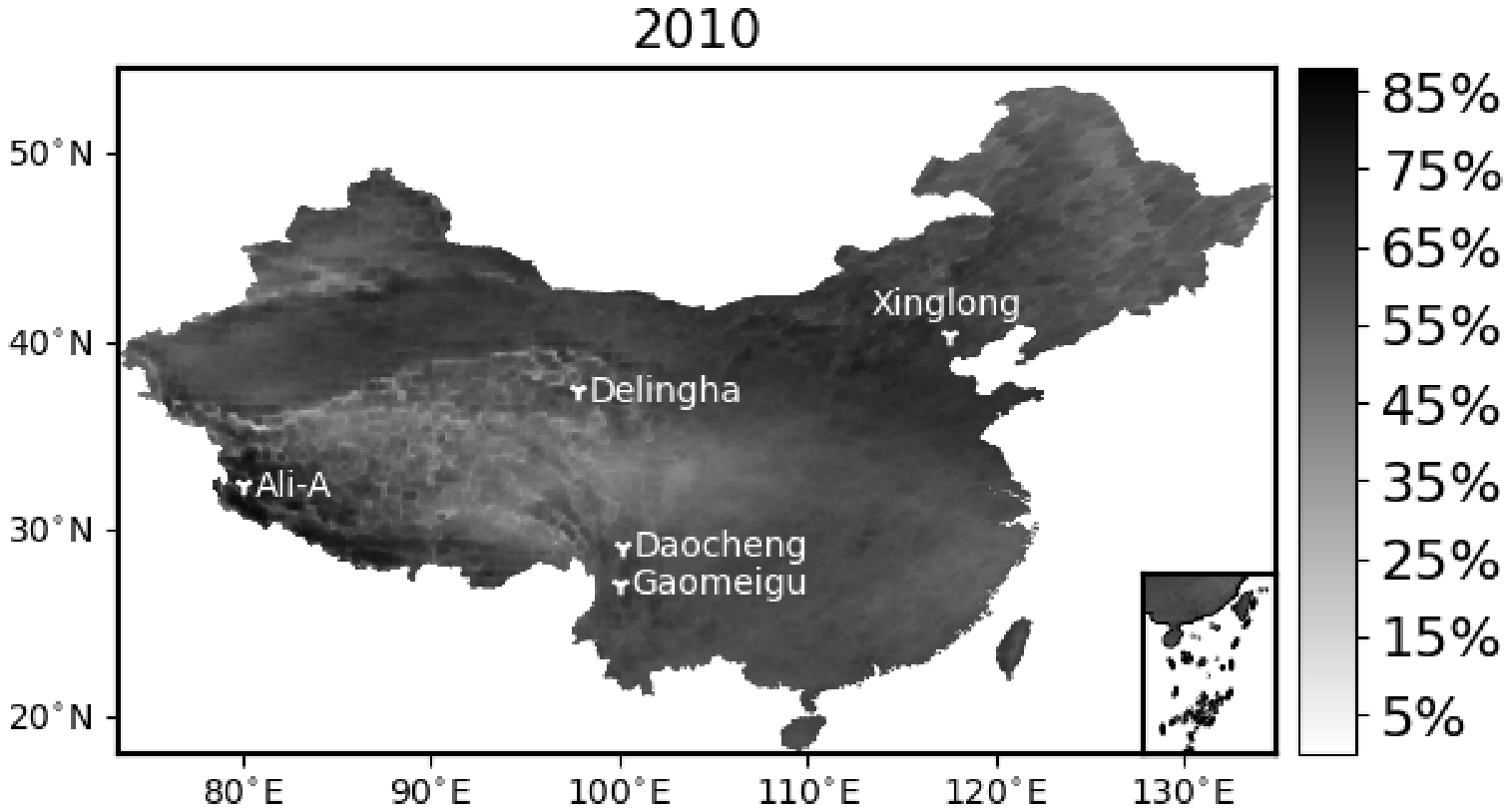}}
{\includegraphics[width=0.32\textwidth]{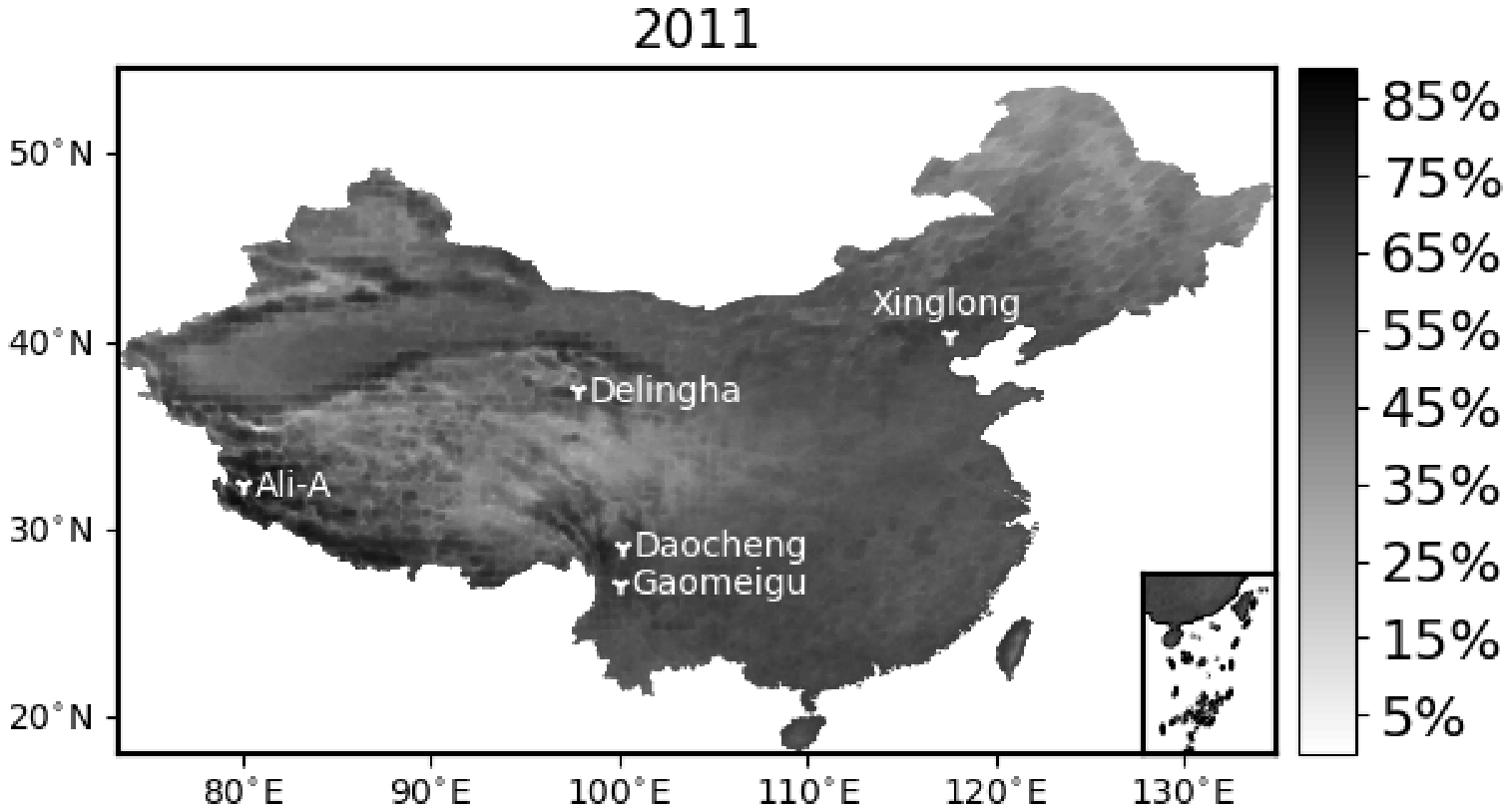}}
{\includegraphics[width=0.32\textwidth]{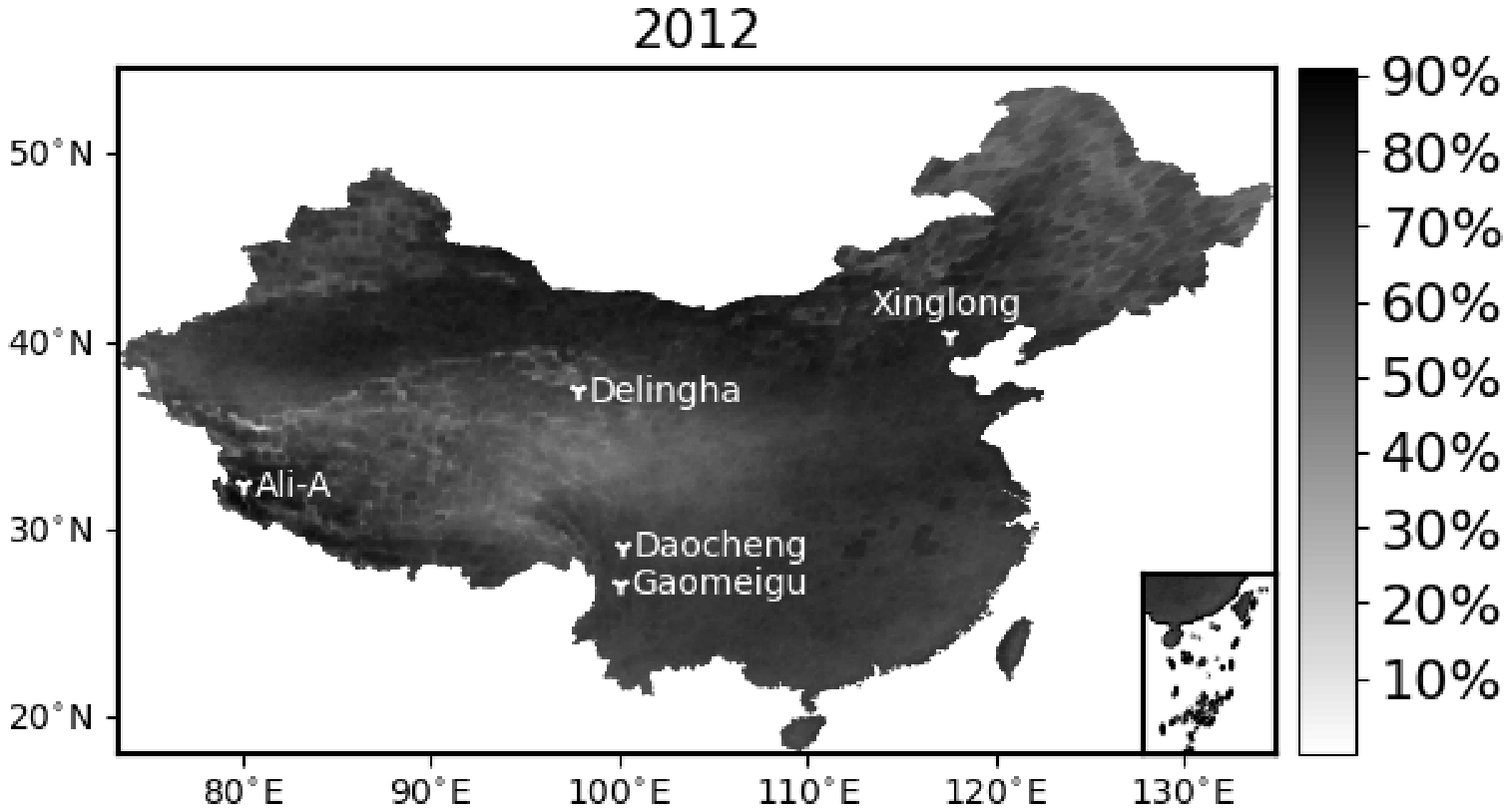}}
{\includegraphics[width=0.32\textwidth]{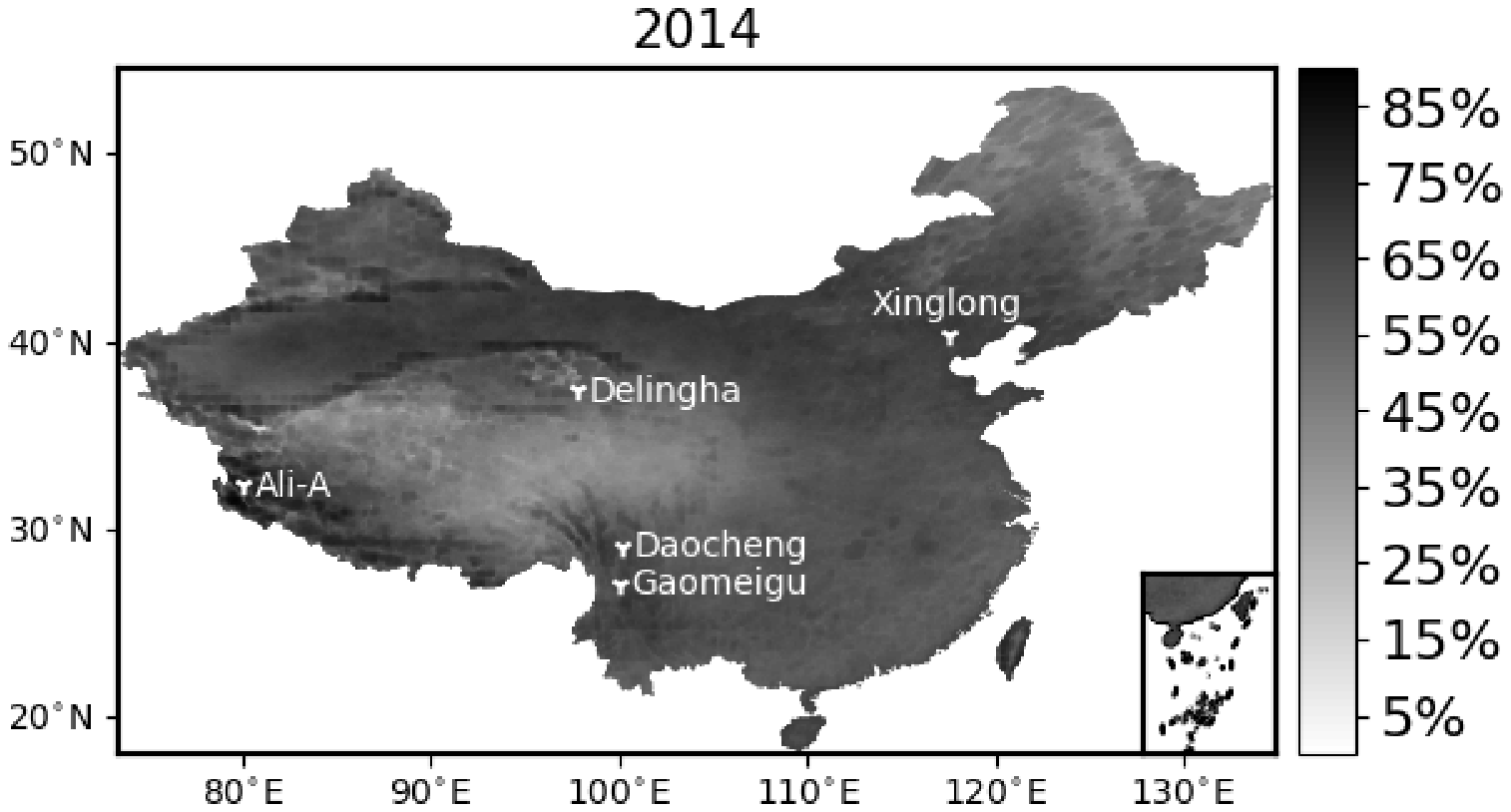}}
{\includegraphics[width=0.32\textwidth]{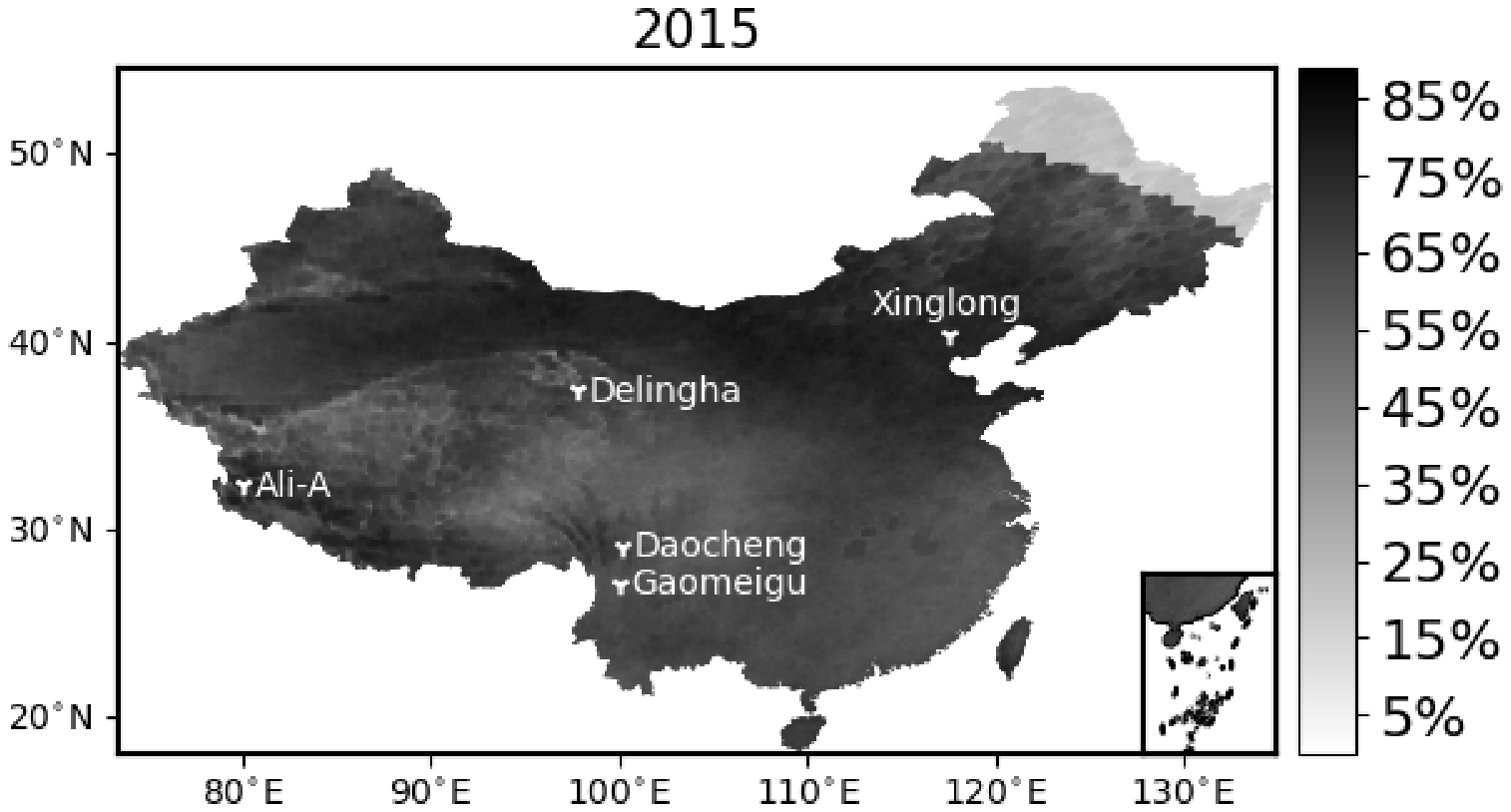}}
{\includegraphics[width=0.32\textwidth]{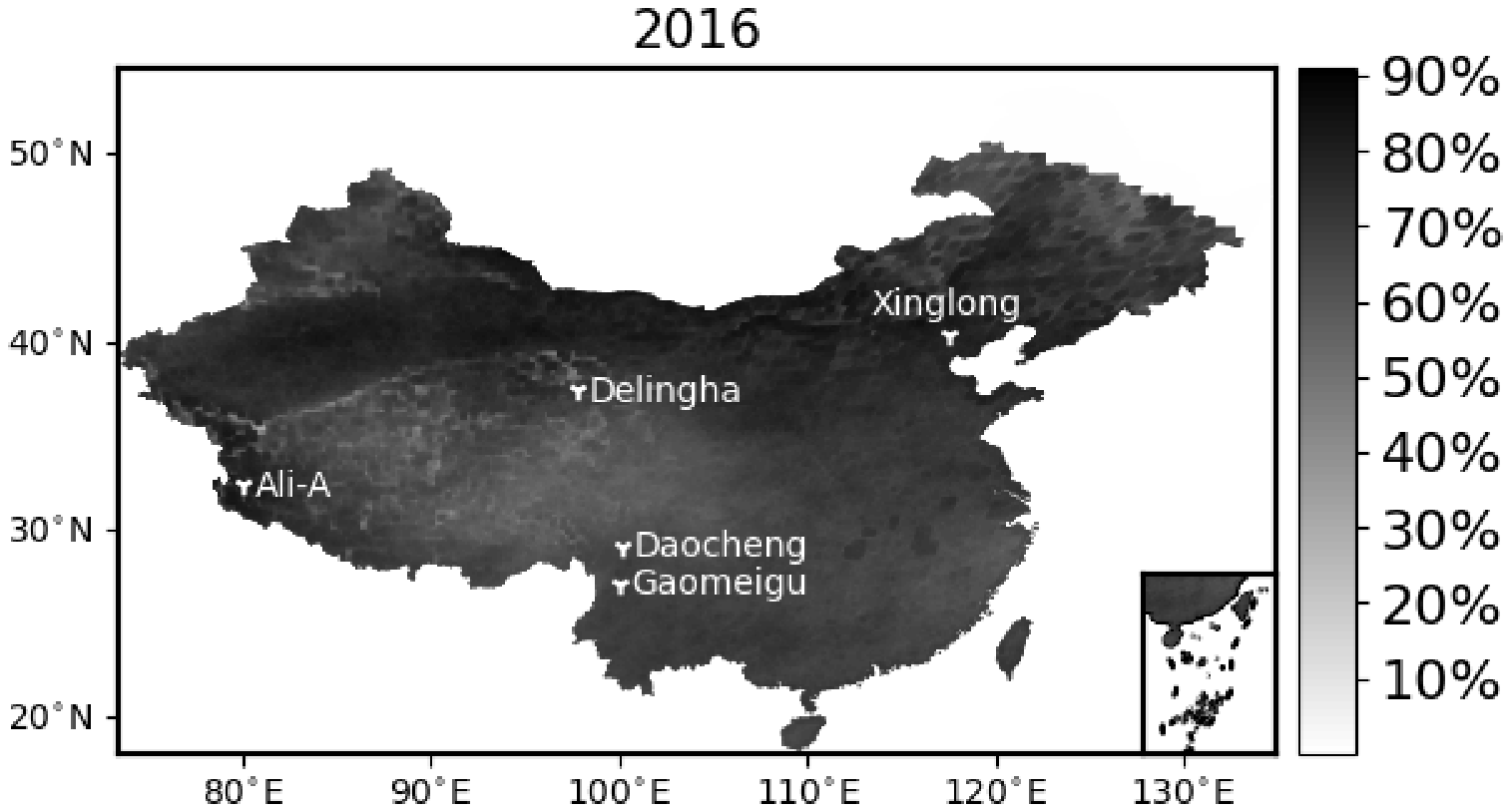}}
{\includegraphics[width=0.32\textwidth]{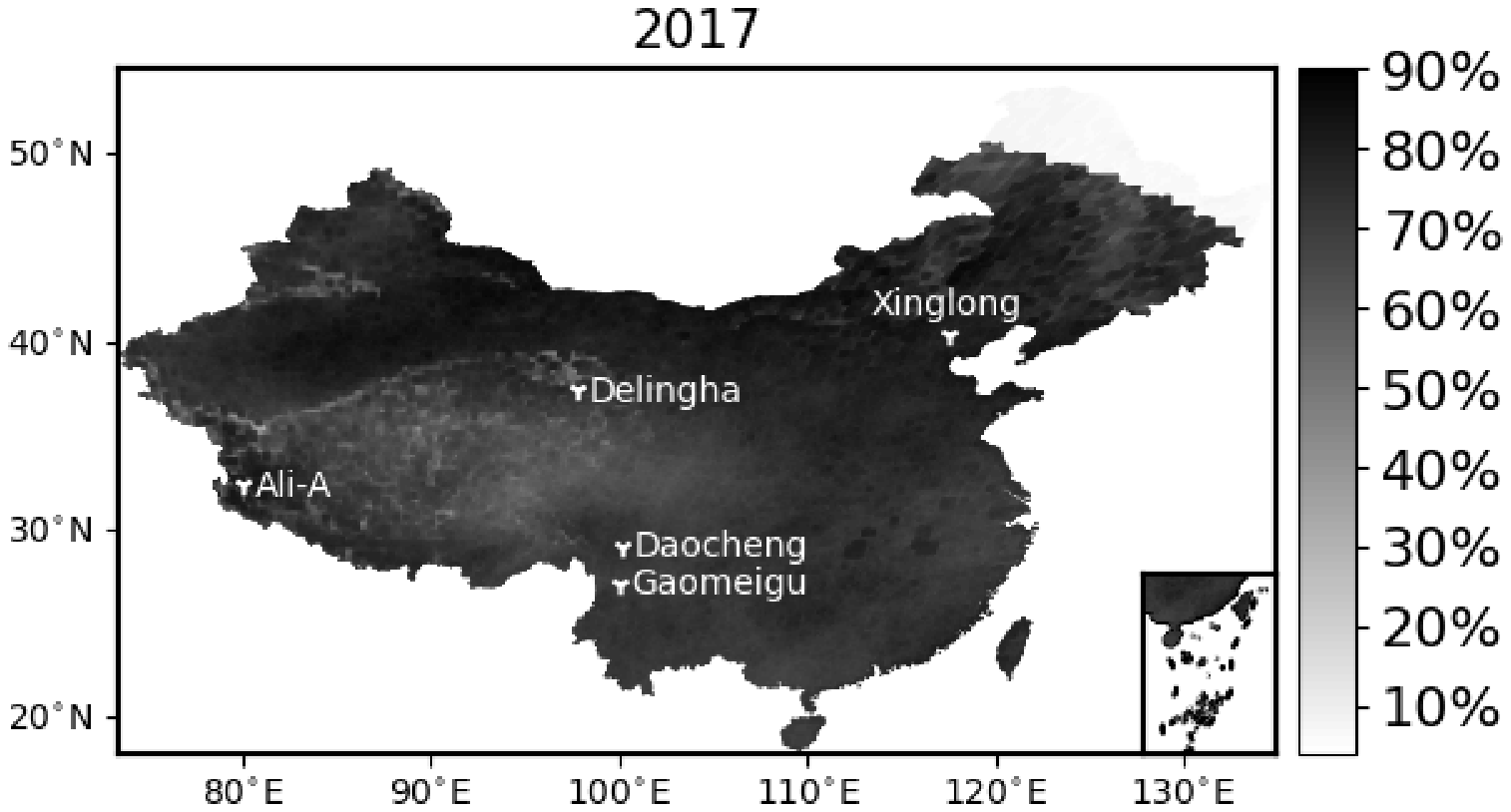}}
{\includegraphics[width=0.32\textwidth]{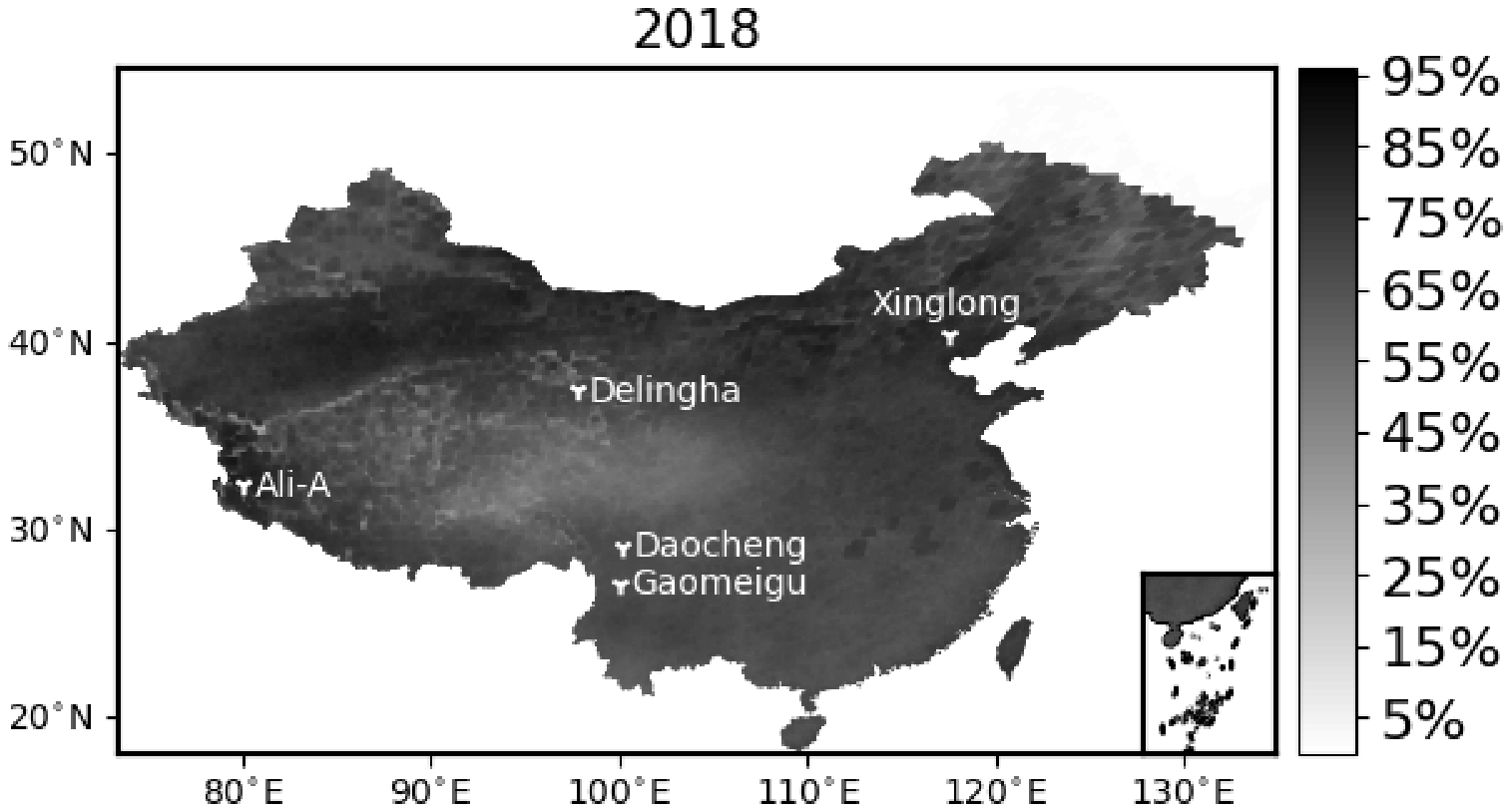}}
{\includegraphics[width=0.32\textwidth]{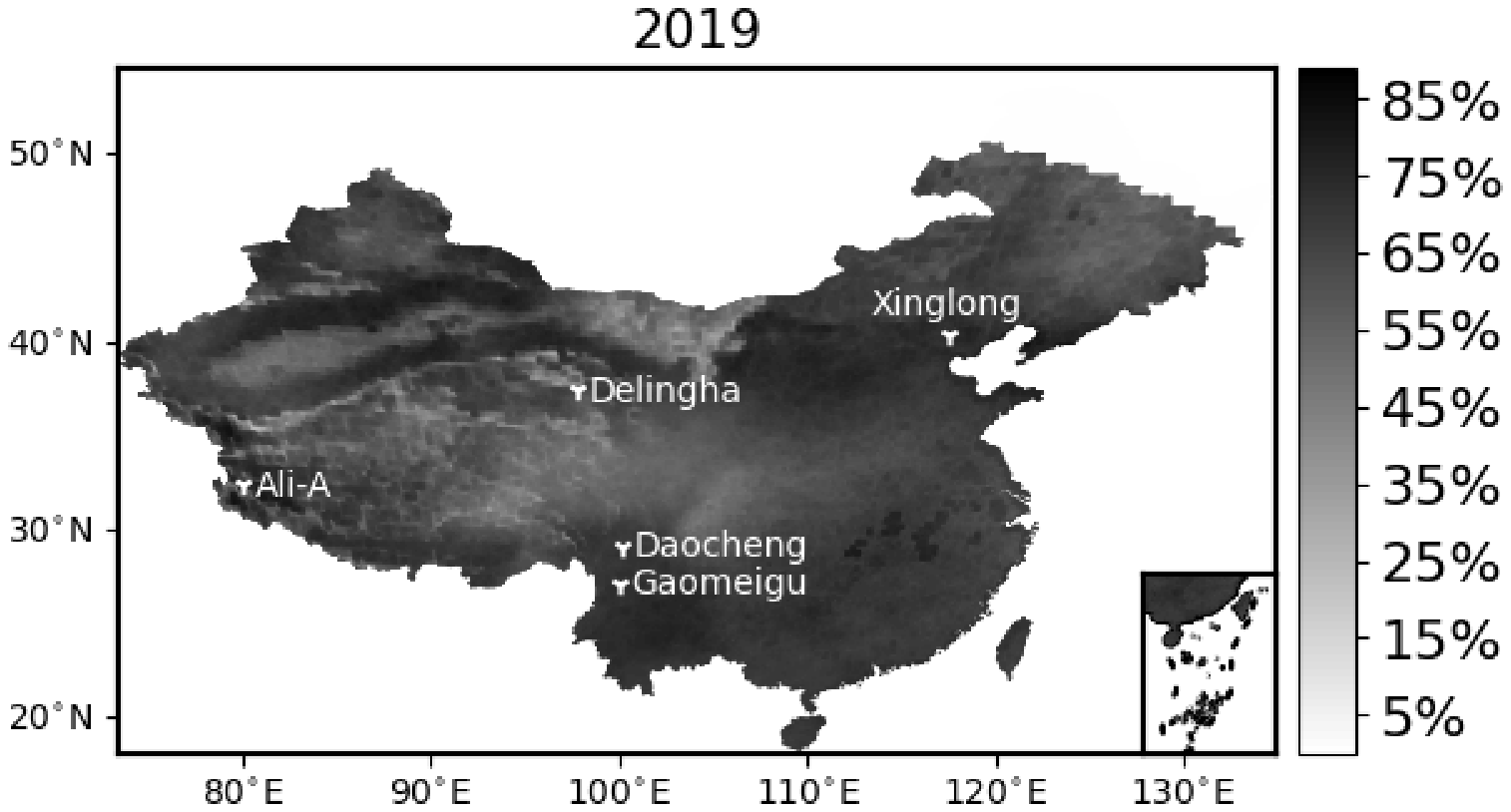}}
{\includegraphics[width=0.7\textwidth]{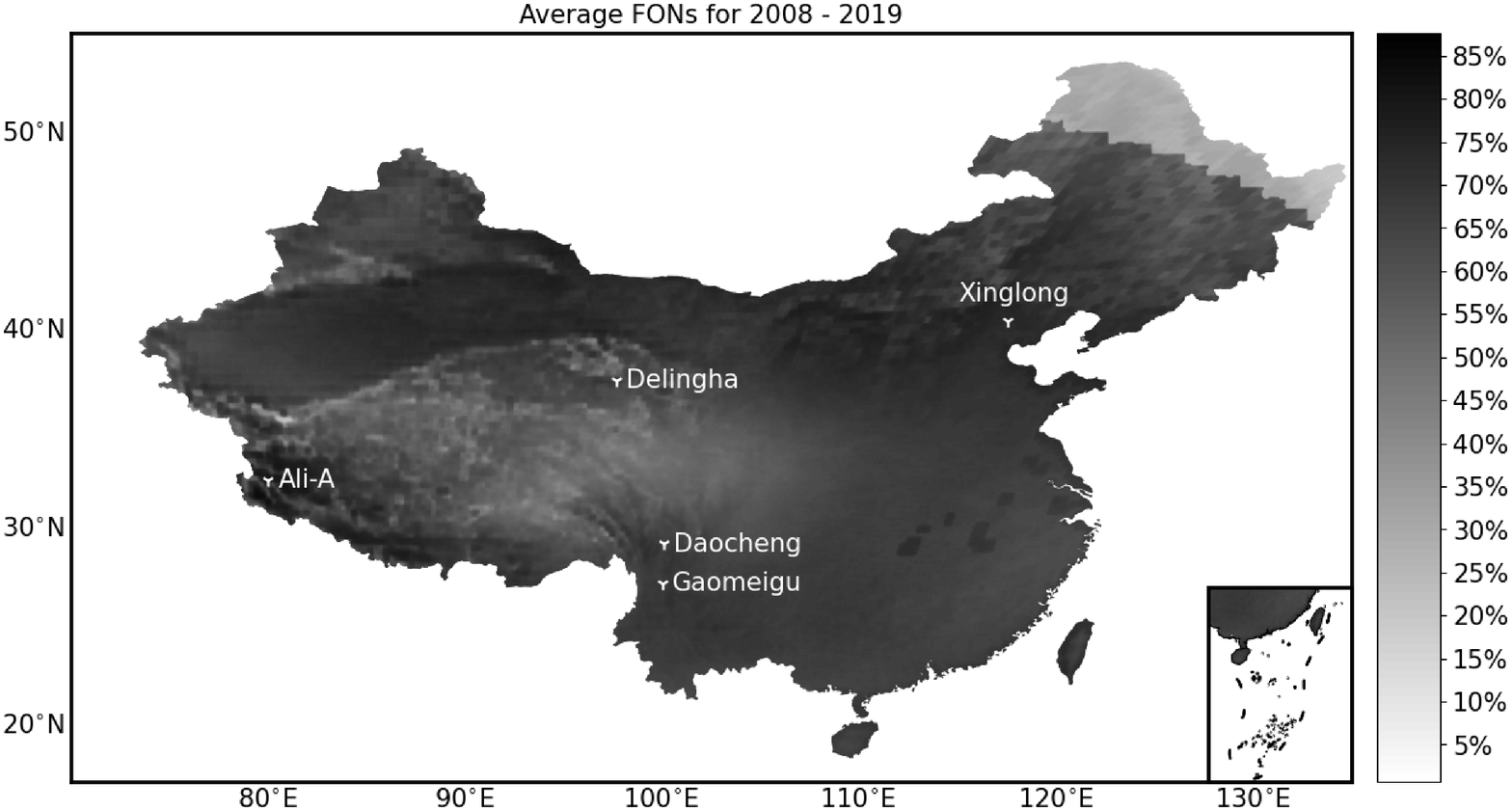}}
\caption{Yearly average FONs in China for 2008 to 2019, except for 2013. The locations of the astronomical sites discussed in the main text are marked as inverted, white triangles.}
\label{ChinaFONs}
\end{figure*}

According to our calculations, sites in Western China, such as Ali-A and Daocheng, show higher FONs than in East China. Among Ali-A, Daocheng, Delingha, and Xinglong, Ali-A has the highest FONs, which is viewed as one of the best astronomical sites in China. The yearly geographical distribution of the FONs above China for 2008 - 2019, except for 2013, and the average of them are shown in Figure~\ref{ChinaFONs}, which indicates most areas in southern China have lower FONs than North and Western China. Moreover, due to the high altitude of the Yun--Gui plateau, Daocheng is endowed with good observing conditions, which provides FONs of more than 70$\%$. North China has better sky conditions for astronomical observations, which has resulted in observatories being built therein (e.g., Xinglong Station).

\begin{table}
\caption{Yearly FONs of Ali-A and Daocheng.}
\label{tab: diff}
\centering
\begin{tabular}{p{7mm}<{\centering}p{7mm}<{\centering}p{15mm}<{\centering}p{7mm}<{\centering}p{4mm}<{\centering}p{15mm}<{\centering}p{7mm}<{\centering}}
\hline
     & \multicolumn{3}{c}{ Ali-A  } & \multicolumn{3}{c}{Daocheng} \\
\hline
Longitude& \multicolumn{3}{c}{  $80^{\circ}01^{\prime}36.16\arcsec$ } & \multicolumn{3}{c}{$100^{\circ}06^{\prime}32.04\arcsec$} \\     
Latitude& \multicolumn{3}{c}{$32^{\circ}19^{\prime}32.63\arcsec$} & \multicolumn{3}{c}{$29^{\circ}06^{\prime}25.02\arcsec$} \\
Altitude& \multicolumn{3}{c}{5040 m } & \multicolumn{3}{c}{4739 m} \\
\hline
Year & C20    & This work   & Diff  & C20      & This work    & Diff    \\
  & ($\%$)    & ($\%$)   & ($\%$)  & ($\%$)      & ($\%$)    & ($\%$)    \\
\hline
2008	&	87.95 	&	84.05 	&	3.90 	&	73.42 	&	70.64 	&	2.78 	\\
2009	&	89.32 	&	84.60 	&	4.72 	&	72.33 	&	72.28 	&	0.05 	\\
2010	&	84.93 	&	81.31 	&	3.62 	&	73.15 	&	72.83 	&	0.32 	\\
2011	&	87.40 	&	76.93 	&	10.47 	&	77.26 	&	66.80 	&	10.46 	\\
2012	&	85.21 	&	78.85 	&	6.36 	&	70.96 	&	69.82 	&	1.14 	\\
2014	&	89.59 	&	88.98 	&	0.61 	&	73.97 	&	75.84 	&	-1.87 	\\
2015	&	85.21 	&	82.14 	&	3.07 	&	70.41 	&	70.09 	&	0.32 	\\
Average	&	87.43 	&	82.41 	&	5.02 	&	73.18 	&	71.19 	&	1.89 	\\
\hline
Year          & F20         & This work         & Diff         & F20          & This work          & Diff            \\
  & ($\%$)    & ($\%$)   & ($\%$)  & ($\%$)      & ($\%$)    & ($\%$)    \\
\hline
2017$^{a}$   & 57.81 & 60.27 & -2.47 & 37.81 & 46.58 & -8.77 \\
2018$^{b}$  & 85.75 & 78.08 & 7.67  & 53.97 & 55.07 & -1.10 \\
Average &71.78	&69.18	&2.6&	45.89&	50.83&	-4.94\\
\hline
\end{tabular}
\begin{flushleft}
\textbf{Note:} `C20' and `F20' mean \cite{Cao2020} and \cite{Feng2020} respectively. `Diff' is the difference between the results from previous works \citep{Cao2020, Feng2020} and this work. It is also worth noting that a) the FONs of 2017 does not include the first three months, because \cite{Feng2020} did not present complete data of them, b) the FONs of 2018 for Daocheng excludes the period of May-July due to the non-operation of the all-sky camera.\\
\end{flushleft}
\end{table}

There are some possible error sources to explain the deviations between our and the literature results. The satellite data of FY-2 have inherent errors. \citet{Xi2013} validated and evaluated the CTA product of FY-2 from 2005 to 2010 and found that the accuracy of cloud detection, missed detected rate, and false detected rate are 82.10$\%$, 6.85$\%$ and 11.05$\%$, respectively. In terms of CTA, a clear sky of low temperature or snow on the ground can be mistakenly detected as clouds. According to \citet{Xu2008}, the maximum error of UTH can be 20$\%$. Moreover, FY-2 cannot detect haze and fog, which adversely affects the estimates of the FONs.

\section{Conclusions}
\label{sec:conclusions}

In this work, we presented a method to calculate the FONs at astronomical sites. The method is based on satellite data (specifically, the CTA and UTH products) of the Chinese Fengyun-2 series of geostationary meteorological satellites, ground data derived from the observation logs of the 60/90-cm Schmidt telescope at Xinglong Station, and the statistics of observable nights at Xinglong Station, Delingha Station and Lijiang Station presented by \citet{zhang2015}, \citet{Tian2016} and \citet{Xin2020}, respectively. To recognise observable nights accurately, we derived optimal CTA and UTH thresholds, finding 35$\%$ and 50$\%$, respectively, based on ground and satellite data.

To validate our method, we calculated the FONs at 27 sites in Indonesia and compared the results with those of \citep{Hidayat2012}, which showed good agreement. We also applied our method to two astronomical sites in China, Ali-A and Daocheng. The average fractions of observable nights of Ali-A and Daocheng derived from our method agree with the results from \citet{Cao2020} and \citet{Feng2020}. Moreover, we calculated the average distribution of the FONs above China from 2008 to 2019. According to our results, the area near 40$^{\circ}$N, on the southern Qinghai--Tibetan plateau, and Yun--Gui plateau, have more observable nights than other areas in China. Among the sites we tested, Ali-A has the most average FONs, 77.75$\%$. 

\section{Acknowledgements}

We thank the anonymous reviewer for insightful and constructive suggestions that greatly strengthened the scientific content of this paper. This work is supported by the China National Astronomical Data Center (NADC), Chinese Virtual Observatory (China-VO), Astronomical Big Data Joint Research Center, co-founded by National Astronomical Observatories, Chinese Academy of Sciences, and Alibaba Cloud.

\section{Data availability}
The FengYun-2 satellite data are publicly available at \url{http://www.nsmc.org.cn/nsmc/en/home/index.html}. The derived data generated in this research will be shared on reasonable request to the corresponding author.

\bibliographystyle{mnras}
\bibliography{sample63} 

\begin{thebibliography}{}
\makeatletter
\relax
\def\mn@urlcharsother{\let\do\@makeother \do\$\do\&\do\#\do\^\do\_\do\%\do\~}
\def\mn@doi{\begingroup\mn@urlcharsother \@ifnextchar [ {\mn@doi@}
  {\mn@doi@[]}}
\def\mn@doi@[#1]#2{\def\@tempa{#1}\ifx\@tempa\@empty \href
  {http://dx.doi.org/#2} {doi:#2}\else \href {http://dx.doi.org/#2} {#1}\fi
  \endgroup}
\def\mn@eprint#1#2{\mn@eprint@#1:#2::\@nil}
\def\mn@eprint@arXiv#1{\href {http://arxiv.org/abs/#1} {{\tt arXiv:#1}}}
\def\mn@eprint@dblp#1{\href {http://dblp.uni-trier.de/rec/bibtex/#1.xml}
  {dblp:#1}}
\def\mn@eprint@#1:#2:#3:#4\@nil{\def\@tempa {#1}\def\@tempb {#2}\def\@tempc
  {#3}\ifx \@tempc \@empty \let \@tempc \@tempb \let \@tempb \@tempa \fi \ifx
  \@tempb \@empty \def\@tempb {arXiv}\fi \@ifundefined
  {mn@eprint@\@tempb}{\@tempb:\@tempc}{\expandafter \expandafter \csname
  mn@eprint@\@tempb\endcsname \expandafter{\@tempc}}}

\bibitem[\protect\citeauthoryear{{Aksaker} et~al.,}{{Aksaker}
  et~al.}{2015}]{Aksaker2015}
{Aksaker} N.,  et~al., 2015, \mn@doi [Experimental Astronomy]
  {10.1007/s10686-015-9458-x}, \href
  {https://ui.adsabs.harvard.edu/abs/2015ExA....39..547A} {39, 547}

\bibitem[\protect\citeauthoryear{{Aksaker}, {Yerli}, {Erdo{\v{g}}an}, {Kurt},
  {Kaba}, {Bayazit}  \& {Yesilyaprak}}{{Aksaker} et~al.}{2020}]{Aksaker2020}
{Aksaker} N.,  {Yerli} S.~K.,  {Erdo{\v{g}}an} M.~A.,  {Kurt} Z.,  {Kaba} K.,
  {Bayazit} M.,   {Yesilyaprak} C.,  2020, \mn@doi [\mnras]
  {10.1093/mnras/staa201}, \href
  {https://ui.adsabs.harvard.edu/abs/2020MNRAS.493.1204A} {493, 1204}

\bibitem[\protect\citeauthoryear{{Cao} et~al.,}{{Cao} et~al.}{2020}]{Cao2020}
{Cao} Z.-H.,  et~al., 2020, \mn@doi [Research in Astronomy and Astrophysics]
  {10.1088/1674-4527/20/6/81}, \href
  {https://ui.adsabs.harvard.edu/abs/2020RAA....20...81C} {20, 081}

\bibitem[\protect\citeauthoryear{{Cavazzani}, {Ortolani}, {Zitelli}  \&
  {Maruccia}}{{Cavazzani} et~al.}{2011}]{Cavazzani2011}
{Cavazzani} S.,  {Ortolani} S.,  {Zitelli} V.,   {Maruccia} Y.,  2011, \mn@doi
  [\mnras] {10.1111/j.1365-2966.2010.17766.x}, \href
  {https://ui.adsabs.harvard.edu/abs/2011MNRAS.411.1271C} {411, 1271}

\bibitem[\protect\citeauthoryear{{Daniyal} \& {Hassan Kazmi}}{{Daniyal} \&
  {Hassan Kazmi}}{2019}]{Daniyal2019}
{Daniyal} {Hassan Kazmi} S.~J.,  2019, \mn@doi [Research in Astronomy and
  Astrophysics] {10.1088/1674-4527/19/9/129}, \href
  {https://ui.adsabs.harvard.edu/abs/2019RAA....19..129D} {19, 129}

\bibitem[\protect\citeauthoryear{{Erasmus} \& {Sarazin}}{{Erasmus} \&
  {Sarazin}}{2001}]{Erasmus2001}
{Erasmus} D.~A.,  {Sarazin} M.~S.,  2001, in {Russell} J.~E.,  {Schaefer} K.,
  {Lado-Bordowsky} O.,  eds,  Society of Photo-Optical Instrumentation
  Engineers (SPIE) Conference Series Vol. 4168, Remote Sensing of Clouds and
  the Atmosphere V. pp 317--328, \mn@doi{10.1117/12.413848}

\bibitem[\protect\citeauthoryear{{Erasmus} \& {Sarazin}}{{Erasmus} \&
  {Sarazin}}{2002}]{Erasmus2002}
{Erasmus} D.,  {Sarazin} M.,  2002, in {Vernin} J.,  {Benkhaldoun} Z.,
  {Mu{\~n}oz-Tu{\~n}{\'o}n} C.,  eds,  Astronomical Society of the Pacific
  Conference Series Vol. 266, Astronomical Site Evaluation in the Visible and
  Radio Range. p.~310

\bibitem[\protect\citeauthoryear{{Erasmus} \& {van Rooyen}}{{Erasmus} \& {van
  Rooyen}}{2006}]{Erasmus2006}
{Erasmus} D.~A.,  {van Rooyen} R.,  2006, {A satellite survey of cloud cover
  and water vapor in northwest Africa and southern Spain}.
p. 62671O, \mn@doi{10.1117/12.669490}

\bibitem[\protect\citeauthoryear{{Feng} et~al.,}{{Feng}
  et~al.}{2020}]{Feng2020}
{Feng} L.,  et~al., 2020, \mn@doi [Research in Astronomy and Astrophysics]
  {10.1088/1674-4527/20/6/80}, \href
  {https://ui.adsabs.harvard.edu/abs/2020RAA....20...80F} {20, 080}

\bibitem[\protect\citeauthoryear{Glickman \& Zenk}{Glickman \&
  Zenk}{2000}]{glickman2000}
Glickman T.~S.,  Zenk W.,  2000, Glossary of meteorology.
American Meteorological Society

\bibitem[\protect\citeauthoryear{{Graham}, {Sarazin}, {Beniston}, {Collet},
  {Hayoz}, {Neun}  \& {Casals}}{{Graham} et~al.}{2005}]{Graham2005}
{Graham} E.,  {Sarazin} M.,  {Beniston} M.,  {Collet} C.,  {Hayoz} M.,  {Neun}
  M.,   {Casals} P.,  2005, \mn@doi [Meteorological Applications]
  {10.1017/S1350482705001520}, \href
  {https://ui.adsabs.harvard.edu/abs/2005MeApp..12...77G} {12, 77}

\bibitem[\protect\citeauthoryear{Hellemeier, Yang, Sarazin  \&
  Hickson}{Hellemeier et~al.}{2018}]{Hellemeier2018}
Hellemeier J.~A.,  Yang R.,  Sarazin M.,   Hickson P.,  2018, \mn@doi [\mnras]
  {10.1093/mnras/sty2982}, 482, 4941

\bibitem[\protect\citeauthoryear{{Hidayat}, {Mahasena}, {Dermawan}, {Hadi},
  {Premadi}  \& {Herdiwijaya}}{{Hidayat} et~al.}{2012}]{Hidayat2012}
{Hidayat} T.,  {Mahasena} P.,  {Dermawan} B.,  {Hadi} T.~W.,  {Premadi} P.~W.,
   {Herdiwijaya} D.,  2012, \mn@doi [\mnras]
  {10.1111/j.1365-2966.2012.22000.x}, \href
  {https://ui.adsabs.harvard.edu/abs/2012MNRAS.427.1903H} {427, 1903}

\bibitem[\protect\citeauthoryear{{Hotan}, {Tingay}  \& {Glazebrook}}{{Hotan}
  et~al.}{2013}]{Hotan2013}
{Hotan} C.~E.,  {Tingay} S.~J.,   {Glazebrook} K.,  2013, \mn@doi [\pasa]
  {10.1017/pasa.2012.002}, \href
  {https://ui.adsabs.harvard.edu/abs/2013PASA...30....2H} {30, e002}

\bibitem[\protect\citeauthoryear{{Lombardi}, {Zitelli}, {Ortolani}  \&
  {Pedani}}{{Lombardi} et~al.}{2006}]{Lombardi2006}
{Lombardi} G.,  {Zitelli} V.,  {Ortolani} S.,   {Pedani} M.,  2006, \mn@doi
  [\pasp] {10.1086/507344}, \href
  {https://ui.adsabs.harvard.edu/abs/2006PASP..118.1198L} {118, 1198}

\bibitem[\protect\citeauthoryear{{Lombardi}, {Zitelli}, {Ortolani}  \&
  {Pedani}}{{Lombardi} et~al.}{2007}]{Lombardi2007}
{Lombardi} G.,  {Zitelli} V.,  {Ortolani} S.,   {Pedani} M.,  2007, \mn@doi
  [\pasp] {10.1086/513079}, \href
  {https://ui.adsabs.harvard.edu/abs/2007PASP..119..292L} {119, 292}

\bibitem[\protect\citeauthoryear{{Lombardi}, {Zitelli}, {Ortolani}, {Pedani}
  \& {Ghedina}}{{Lombardi} et~al.}{2008}]{Lombardi2008}
{Lombardi} G.,  {Zitelli} V.,  {Ortolani} S.,  {Pedani} M.,   {Ghedina} A.,
  2008, \mn@doi [\aap] {10.1051/0004-6361:20078372}, \href
  {https://ui.adsabs.harvard.edu/abs/2008A&A...483..651L} {483, 651}

\bibitem[\protect\citeauthoryear{{Ma} et~al.,}{{Ma} et~al.}{2020}]{Ma2020}
{Ma} B.,  et~al., 2020, \mn@doi [\nat] {10.1038/s41586-020-2489-0}, \href
  {https://ui.adsabs.harvard.edu/abs/2020Natur.583..771M} {583, 771}

\bibitem[\protect\citeauthoryear{{Morrison}, {Murphy}, {Cruikshank}, {Sinton}
  \& {Martin}}{{Morrison} et~al.}{1973}]{Morrison1973}
{Morrison} D.,  {Murphy} R.~E.,  {Cruikshank} D.~P.,  {Sinton} W.~M.,
  {Martin} T.~Z.,  1973, \mn@doi [\pasp] {10.1086/129449}, \href
  {https://ui.adsabs.harvard.edu/abs/1973PASP...85..255M} {85, 255}

\bibitem[\protect\citeauthoryear{{Sarazin}, {Graham}  \&
  {Kurlandczyk}}{{Sarazin} et~al.}{2006}]{Sarazin2006}
{Sarazin} M.,  {Graham} E.,   {Kurlandczyk} H.,  2006, The Messenger, \href
  {https://ui.adsabs.harvard.edu/abs/2006Msngr.125...44S} {125, 44}

\bibitem[\protect\citeauthoryear{{Sch{\"o}ck} et~al.,}{{Sch{\"o}ck}
  et~al.}{2009}]{Schock2009}
{Sch{\"o}ck} M.,  et~al., 2009, \mn@doi [\pasp] {10.1086/599287}, \href
  {https://ui.adsabs.harvard.edu/abs/2009PASP..121..384S} {121, 384}

\bibitem[\protect\citeauthoryear{{Sinnott}}{{Sinnott}}{1984}]{Sinnott1984}
{Sinnott} R.~W.,  1984, \skytel, \href
  {https://ui.adsabs.harvard.edu/abs/1984S&T....68R.158S} {68, 158}

\bibitem[\protect\citeauthoryear{{Site Testing Group of Yunnan Astronomical
  Observatory}}{{Site Testing Group of Yunnan Astronomical
  Observatory}}{1999}]{Site1999}
{Site Testing Group of Yunnan Astronomical Observatory} 1999, Acta Astronomica
  Sinica, 3, 326

\bibitem[\protect\citeauthoryear{{Tapia}}{{Tapia}}{1992}]{Tapia1992}
{Tapia} M.,  1992, \rmxaa, \href
  {https://ui.adsabs.harvard.edu/abs/1992RMxAA..24..179T} {24, 179}

\bibitem[\protect\citeauthoryear{{Tian} et~al.,}{{Tian}
  et~al.}{2016}]{Tian2016}
{Tian} J.~F.,  et~al., 2016, \mn@doi [\pasp]
  {10.1088/1538-3873/128/968/105003}, \href
  {https://ui.adsabs.harvard.edu/abs/2016PASP..128j5003T} {128, 105003}

\bibitem[\protect\citeauthoryear{{Varela}, {Bertolin},
  {Mu{\~n}oz-Tu{\~n}{\'o}n}, {Ortolani}  \& {Fuensalida}}{{Varela}
  et~al.}{2008}]{Varela2008}
{Varela} A.~M.,  {Bertolin} C.,  {Mu{\~n}oz-Tu{\~n}{\'o}n} C.,  {Ortolani} S.,
   {Fuensalida} J.~J.,  2008, \mn@doi [\mnras]
  {10.1111/j.1365-2966.2008.13803.x}, \href
  {https://ui.adsabs.harvard.edu/abs/2008MNRAS.391..507V} {391, 507}

\bibitem[\protect\citeauthoryear{{Vernin}, {Mu{\~n}oz-Tu{\~n}on}  \&
  {Sarazin}}{{Vernin} et~al.}{2008}]{Vernin2008}
{Vernin} J.,  {Mu{\~n}oz-Tu{\~n}on} C.,   {Sarazin} M.,  2008, in {Stepp}
  L.~M.,  {Gilmozzi} R.,  eds,  Society of Photo-Optical Instrumentation
  Engineers (SPIE) Conference Series Vol. 7012, Ground-based and Airborne
  Telescopes II. p. 70121T, \mn@doi{10.1117/12.788731}

\bibitem[\protect\citeauthoryear{{Xi}, {Shi}, {Zhao}, {Zhu}  \& {Huang}}{{Xi}
  et~al.}{2013}]{Xi2013}
{Xi} L.,  {Shi} C.-X.,  {Zhao} B.-F.,  {Zhu} C.,   {Huang} X.-L.,  2013,
  Meteorogical Science and Technology, 41, 8

\bibitem[\protect\citeauthoryear{{Xin} et~al.,}{{Xin} et~al.}{2020}]{Xin2020}
{Xin} Y.-X.,  et~al., 2020, \mn@doi [Research in Astronomy and Astrophysics]
  {10.1088/1674-4527/20/9/149}, \href
  {https://ui.adsabs.harvard.edu/abs/2020RAA....20..149X} {20, 149}

\bibitem[\protect\citeauthoryear{{Xu}, {Zhang}, {Yang}  \& {Zhao}}{{Xu}
  et~al.}{2008}]{Xu2008}
{Xu} J.,  {Zhang} W.,  {Yang} J.,   {Zhao} L.,  2008, A practical manual on
  operational products and satellite data formats for Fengyun-2 satellites.
China Meteorological Press, Beijing, China

\bibitem[\protect\citeauthoryear{{Yao}}{{Yao}}{2005}]{yao2005}
{Yao} Y.-Q.,  2005, Journal of The Korean Astronomical Society, 38, 113

\bibitem[\protect\citeauthoryear{{Zhang}, {Ge}, {Lu}, {Cao}, {Chen}, {Mao}  \&
  {Jiang}}{{Zhang} et~al.}{2015}]{zhang2015}
{Zhang} J.-C.,  {Ge} L.,  {Lu} X.-M.,  {Cao} Z.-H.,  {Chen} X.,  {Mao} Y.-N.,
  {Jiang} X.-J.,  2015, \mn@doi [\pasp] {10.1086/684369}, \href
  {https://ui.adsabs.harvard.edu/abs/2015PASP..127.1292Z} {127, 1292}

\makeatother
\end{thebibliography}
\end{document}